\documentclass[fleqn,usenatbib]{mnras}

\usepackage{newtxtext,newtxmath}

\usepackage[T1]{fontenc}

\DeclareRobustCommand{\VAN}[3]{#2}
\let\VANthebibliography\thebibliography
\def\thebibliography{\DeclareRobustCommand{\VAN}[3]{##3}\VANthebibliography}

\usepackage{graphicx}	
\usepackage{amsmath}	
\usepackage{xcolor}
\usepackage{multirow}   
\usepackage{multicol}
\usepackage{booktabs}   
\usepackage{soul}

\newcommand{\mr}[1]{\mathrm{#1}}

\title[MIGHTEE-POL Data Release]
{MIGHTEE Polarization Early Science Fields: The Deep Polarized Sky}

\author[A. R. Taylor et al.]{
A. R. Taylor$^{1,2,3}$\thanks{E-mail: russ@idia.ac.za},
S. Sekhar$^{1,2,3,4}$,
L. Heino$^{1,2}$,
A.~M.~M. Scaife$^{5,6}$,
J. Stil$^{7}$,
M. Bowles$^{5}$,
M. Jarvis$^{8}$, 
\newauthor
I. Heywood$^{8,9,10}$,
J. D. Collier$^{1,2,11}$
\\
$^1$Inter-University Institute for Data Intensive Astronomy, Cape Town, 
South Africa\\
$^2$ Department of Astronomy, University of Cape Town, South Africa \\
$^3$ Department of Physics and Astronomy, University of the Western Cape, South Africa\\
$^4$ National Radio Astronomy Observatory, Socorro, New Mexico, USA \\
$^5$Department of Physics and Astronomy, University of Manchester, Manchester, UK\\
$^6$The Alan Turing Institute, Euston Road, London, NW1 2DB, UK\\
$^7$Department of Physics and Astronomy, University of Calgary, Calgary, Canada\\
$^8$Department of Astrophysics, University of Oxford, Oxford, UK\\
$^9$Centre for Radio Astronomy Techniques and Technologies, Department of Physics and Electronics, Rhodes University, Makhanda, South Africa\\
$^{10}$ South African Radio Astronomy Observatory, Cape Town, South Africa\\
$^{11}$ School of Science, Western Sydney University, Locked Bag 1797, Penrith, NSW 2751, Australia\\
}

\pubyear{2023}

\begin{document}
\label{firstpage}
\pagerange{\pageref{firstpage}--\pageref{lastpage}}
\maketitle

\begin{abstract}
 The MeerKAT International GigaHertz Tiered Extragalactic Exploration (MIGHTEE) is one of the MeerKAT large survey projects, designed to pathfind SKA key science. MIGHTEE is undertaking deep radio imaging of four
well observed fields (COSMOS, XMM-LSS, ELAIS\,S1 and CDFS) totaling 20 square degrees to $\mu$Jy sensitivities. Broadband imaging observations between 
880--1690 MHz yield total intensity continuum, spectro-polarimetry, and atomic hydrogen spectral imaging.
Early science data from MIGHTEE are being released from initial observations of COSMOS and XMM-LSS. This paper describes the spectro-polarimetric observations, the polarization data processing of the MIGHTEE early science fields, and presents polarization data images and catalogues.
The catalogues include radio spectral index, redshift information and Faraday rotation measure synthesis results for 13,271 total intensity radio sources down to a polarized intensity detection limit of $\sim$20\,$\mu$Jy\,bm$^{-1}$. 
Polarized signals were detected from 324 sources. 
For the polarized detections we include a catalogue of Faraday Depth from 
both Faraday Synthesis and $Q$, $U$ fitting, as well as total 
intensity and polarization spectral indices.  
The distribution of redshift of the total radio sources and detected polarized sources are the same, with median redshifts of 0.86 and 0.82 respectively.  
Depolarization of the emission at longer-wavelengths is seen to increase with decreasing total-intensity spectral index, implying that depolarisation is intrinsic to the radio sources.
No evidence is seen for a redshift dependence of the variance of
Faraday Depth.

\end{abstract}




\section{Introduction}

The MeerKAT telescope is the precursor of the Square Kilometre Array mid-frequency dish array to begin deployment later this decade on the Karroo Plateau in South Africa.
The MeerKAT MIGHTEE survey \citep{Jarvis_2016} is undertaking deep radio
imaging of four extragalactic fields, COSMOS, XMM-LSS, 
CDFS, and ELAIS\,S1.  All fields are being observed at L-band from 880--1680 MHz,
in multiple pointings that will be mosaiced to a final image with a broadband sensitivity of $\sim$2\,$\mu$Jy\,beam$^{-1}$.  
The MIGHTEE project is creating
science data products for total intensity, broadband continuum science, 
HI spectral line science, and spectro-polarimetric science, to explore dark matter, large scale structure, and the evolution of galaxies.

Radio polarimetry is a powerful probe of cosmic magnetic fields, and is central 
to science programs on next generation radio facilities and to planning for the Square Kilometre Array \citep{Heald_2020}.
Deep polarization imaging offers the potential to explore cosmic magnetism to
flux density levels where the radio population transitions from AGN dominated
to star formation galaxies at intermediate redshifts 
\citep{Taylor_2015}.

The MIGHTEE project is releasing data for early science from initial observations of portions of the 
COSMOS and XMM-LSS regions. 
Early Science total intensity images and catalogues were released by \cite{Heywood_2022}.
This document describes the spectro-polarimetric data processing and early 
science polarisation data products.

\section{Observations}
MeerKAT is an array of 64, 13.5-meter diameter radio antennas located in the Karroo Plateau in South Africa \citep{Jonas_2016}.  The MIGHTEE early science observation consist of one pointing 
centred on the COSMOS field and three adjacent overlapping pointings in XMM-LSS labelled XMMLSS\_12, XMMLSS\_13, XMMLSS\_14.
Each of the pointing centres for the XMM-LSS were observed twice with MeerKAT
for approximately 8 hours.  For the COSMOS pointing three 8-hour observations were
obtained.  Table~\ref{tab:obs} summarizes the observations.  The number of antennas participating in an observation ranged from 
a minimum of 59 to a maximum of the full array of 64.

For the XMMLSS pointings J1939-6342 was used as the primary calibrator for calibration
of the flux density scale and complex bandpass. Time-dependent gain was tracked
by observation of the secondary calibrator J0201-1132 every 20 to 30 minutes during the 8 hour track. The polarized source J0521+1638 (3C138) was used for absolute polarization angle and observed once or twice during a track.
For COSMOS the primary was J0408-6565, the secondary was J1008+0740, and J1331+3030 (3C286)
was used for polarization angle.  Observations taken in 2018 used the 4096 channel
coreeeeeaaarelator mode.  The observation of COSMOS in 2020 was taken with 32768 channels. In that case the visibility data were averaged to 4096 channels before processing.

\begin{table}
    \centering
    \begin{tabular}{|c|c|c|c|}
    \hline
          Pointing & Date & time (h) & N antennas \\
    \hline
          XMMLSS\_12 & 2018-10-06 & 8.02 & 59 \\
          XMMLSS\_12 & 2018-10-11 & 8.05 & 63 \\
          XMMLSS\_13 & 2018-10-07 & 8.07 & 59 \\
          XMMLSS\_13 & 2018-10-12 & 8.03 & 62 \\
          XMMLSS\_14 & 2018-10-08 & 8.03 & 60 \\
          XMMLSS\_14 & 2018-10-13 & 8.00 & 62 \\
          COSMOS &  2018-04-19 & 8.65 &  64 \\
          COSMOS & 2018-05-06 & 8.39 & 62   \\
          COSMOS & 2020-04-26 & 7.98 & 59 \\
    \hline
    \end{tabular}
    \caption{Observations for the MIGHTEE-pol Early Science Release.}
    \label{tab:obs}
\end{table}

\section{Data Processing}
\begin{figure}
    \centering
    \includegraphics[width=\linewidth]{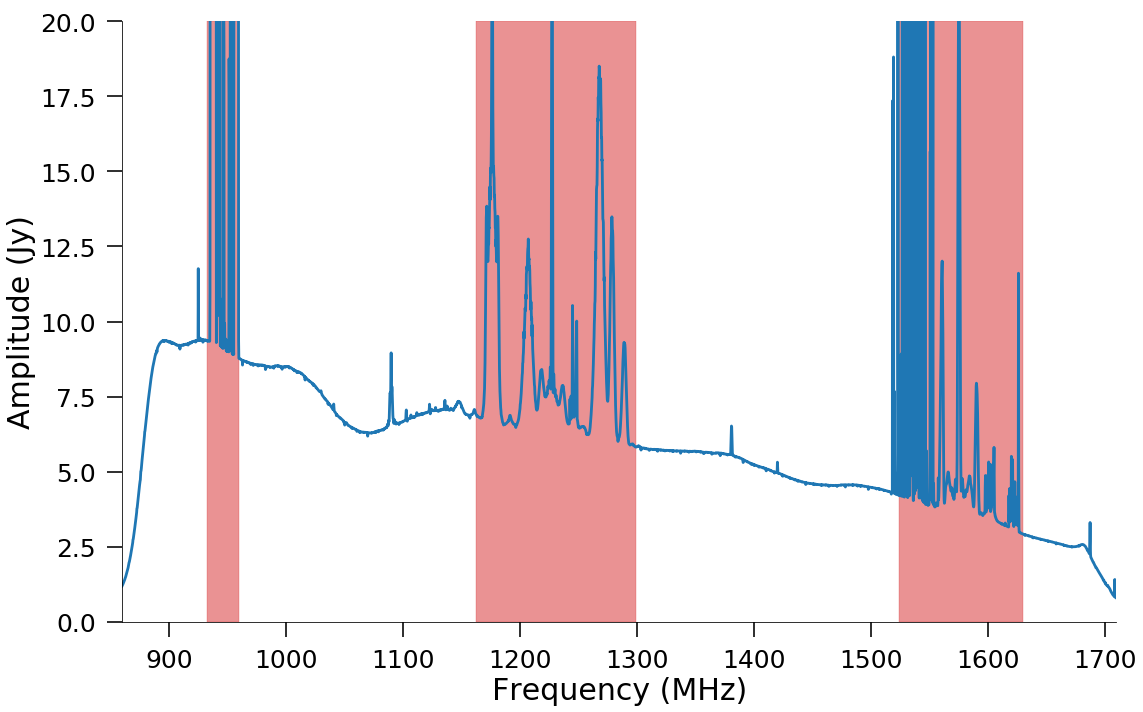}
    \caption{Plot of the MeerKAT time-averaged bandpass response on J1939-6342. The regions highlighted in red have strong, persistent RFI in all observations and are removed
    for all data sets.}
    \label{fig:badfreqranges}
\end{figure}

The visibility data were calibrated and imaged in full polarization mode on the ilifu cloud facility using the \textsc{CASA}-based IDIA pipeline\footnote{\url{https://idia-pipelines.github.io/docs/processMeerKAT}}.
The pipeline partitions the L-band RF into 15 spectral windows between 880\,MHz and 1680\,MHz.  
Data in frequency ranges with strong persistent RFI (933--960\,MHz, 1163--1299\,MHz, and 1525--1630\,MHz) are removed at this partition stage.  
Figure~\ref{fig:badfreqranges} shows a sample plot of the time averaged visibility 
amplitudes on J1939-6342, and illustrates the regions removed due to RFI. 

Each of the 14 spectral windows
is split into its own multi-measurement set (MMS) that is processed concurrently using the 
ilifu SLURM job manager.  
Following calibration the calibrated data from each spectral
window is merged into a single measurement set (MS) with calibrated visibilities.

For ideal linear feeds, the linear approximation to the response of the parallel and cross hand visibilities is given by the equations 1-4 below.  
In these equations we have retained only the first order leakage terms (\textit{i.e.,} those leakage terms that multiply Stokes $I$). A more complete set including leakage terms that multiply the source polarization can be
found in \cite{Hales_2017}.
\begin{eqnarray}
V_{xx} = g^i_{x}g^k_{x} \bigl(I + Q\cos2\psi + U\sin 2\psi \bigr) \\
V_{xy} = g^i_{x}g^k_{y}\bigl [ (d^i_x - d^k_y\ ^*)I - Q \sin 2\psi + U\cos 2\psi + jV\bigr ] \\
V_{yx} = g^i_{y}g^k_{x}\bigl [ (d^k_x\ ^* - d^i_y)I - Q \sin 2\psi + U\cos 2\psi - jV \bigr ] \\
V_{yy} = g^i_{y}g^k_{y} \bigl(I - Q\cos2\psi - U\sin 2\psi \bigr) 
\label{eq:stokes_to_corr}
\end{eqnarray}
Any polarized signal from the source is present in all four correlations and varies with parallactic angle $\psi$. Consequently, the source polarization signal affects both the gain solutions and the leakage. 
The accuracy of the Stokes $Q$ response ($V_{xx}-V_{yy}$) is dependent on the knowledge of the 
complex gains $g_x$ and $g_y$. 
Precise measurement of the gains and the leakage terms can be 
achieved using a strong unpolarized source for which the equations reduce to 
\begin{eqnarray}
V_{xx} = g^i_{x}g^k_{x} I \\
V_{xy} = g^i_{x}g^k_{y} (d^i_x - d^k_y\ ^*)I  \\
V_{yx} = g^i_{y}g^k_{x} (d^k_x\ ^* - d^i_y)I  \\
V_{yy} = g^i_{y}g^k_{y} I
\end{eqnarray}
These equations cleanly separate the gain solution into the parallel-hand correlations and the gain leakage into 
the cross-hand correlations. 

Prior to calibration, the data are first flagged to eliminate the frequency windows that contain persistent RFI, as shown in Fig.~\ref{fig:badfreqranges}.

Following automated flagging, as a first step in the calibration we use the observations of the primary calibrator to measure the frequency-dependent $g_x$ and $g_y$ gains using an 
absolute flux bandpass solution. Assuming the primary calibrators are
unpolarized (but see section~\ref{sec:leakage}), we apply the frequency-dependent
gain solutions and derive the leakage terms also using the primary.

The secondary calibrator is used to calculate polarisation-independent, but
time-dependent, residual complex gains. The XX and YY signals were averaged 
(\texttt{gaintype ='T'} in \textsc{CASA} task \texttt{gaincal}) so any polarisation of the secondary does not modify the relative X, Y gains 
derived from the primary.    
Finally, the scan on the polarization calibrator (one of either J0521+1638 or J1331+3030) was used for absolute X-Y phase calibration. We find that the XY phase solutions are very consistent across all the early science fields, so we fit the median XY phase spectrum with a 3rd degree polynomial and apply the fitted calibration solutions. This results in more stable and smoother Q \& U spectra on the calibrator and target sources.

This first round of calibration is followed by a second round of automated flagging, using the \texttt{tfcrop} and \texttt{rflag} algorithms within the \textsc{CASA} \texttt{flagdata} task. The flagging is carried out on the calibrated data in order to identify low-level RFI in the data. The calibration solutions are then cleared out and the entire calibration process is run again, as described above, resulting in the final calibrated measurement sets. 

Following \textit{a priori} calibration, the calibrated visibilities are self-calibrated
(again using \texttt{gaintype = 'T'} to preserve the relative X,Y gains).
Two self-calibration cycles are run, first calibrating only for phase, followed by a phase and amplitude solve. The self-calibrated target visibilities and calibrator visibilities are moved to the mightee-pol visibility repository on ilifu, along with all calibration tables.  

\begin{figure}
    \centering
    \includegraphics[width=\columnwidth]{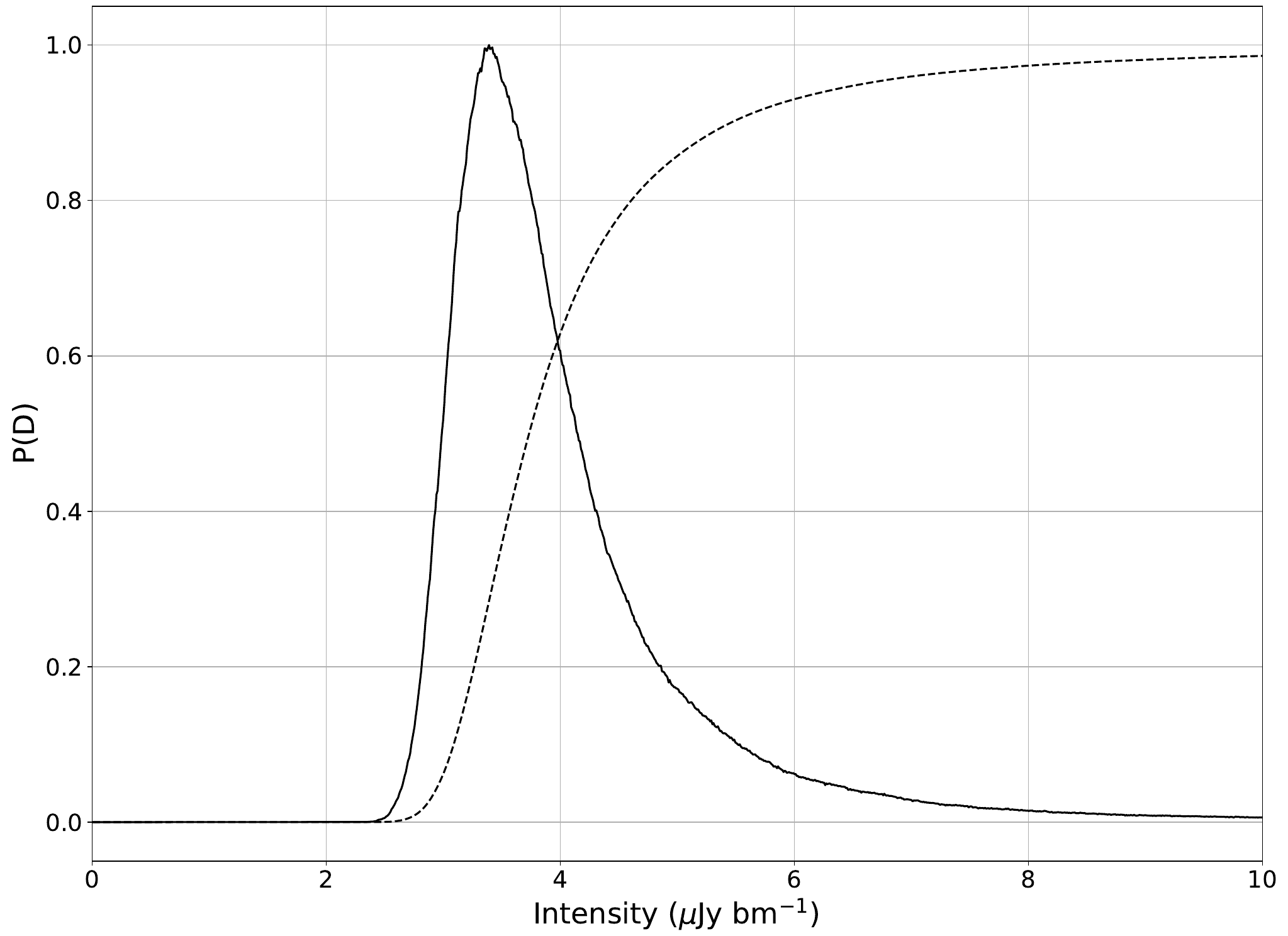}
    \caption{Distribution of pixel amplitudes (solid line) in the pyBDSF rms image for XMMLSS\_12.
    The most probable value (distribution peak) is 3.15\,$\mu$Jy\,bm$^{-1}$. The
    median value is 3.49\,$\mu$Jy\,bm$^{-1}$. The dashed line shows the cumulative distribution.}
    \label{fig:rms}
\end{figure}

Spectro-polarimetric $(I,Q,U,V)$ hypercubes are created from the calibrated visibilities for the calibrators and the self-calibrated visibilities of the target fields using the IDIA cube generation 
pipeline\footnote{\url{https://github.com/idia-astro/frocc}}. 
In addition multi-term multi-frequency synthesis (MT-MFS, \citealt{rau2011multi}) broadband images are created in
full Stokes.
All images were constructed for each field using \textsc{casa} \texttt{tclean} with 
the combined visibilities from the multiple observations. 
Image products created for each field include:
\begin{itemize}
    \item Two MT-MFS broadband $I,Q,U,V$ images with Briggs' robust weighting \citep{briggs1995high} of -0.5 and +0.4. The central frequency of the images is 1.280 GHz.
    \item Spectro-polarimetric $I,Q,U,V$ hypercubes with a Briggs' robust weighting of +0.0. Two hypercubes are constructed. For one, each frequency plane has the native resolution of the associated frequency. 
    A second hypercube is generated with each frequency channel smoothed in the image plane to a common resolution of $18''$. 
\end{itemize} 
For all image data the clean deconvolution is carried out with images of the entire primary beam with frame dimensions of $6144\times6144$ pixels and $1.5''$ cell size.
For science use, the frames are reduced to $4600\times4600$ after deconvolution, 
yielding an image diameter of 1.92$^{\circ}$.
Table~\ref{tab:images} lists the central coordinates of the four image sets and the 
restoring beams of the MFS images at robust $= -0.5$.   The rms noise of the Stokes $I$ 
plane of the MFS images, measured in a source free region far from the image centre is typically
between 3--3.5\,$\mu$Jy\,bm$^{-1}$.
As an example,
Figure~\ref{fig:rms} shows the distribution of pixel amplitudes in the rms image for XMMLSS\_12 created with pyBDSF. 
The most probable value of the rms is 3.15\,$\mu$Jy\,bm$^{-1}$ and the median value is 3.49\,$\mu$Jy\,bm$^{-1}$.

The cubes span a frequency range of 794 MHz (886--1680 MHz) with channel width of 2.51 MHz.  
Channel images are constructed using all frequency channels
within each 2.51 MHz band with \textsc{CASA} \texttt{tclean} in 
multi-frequency synthesis mode 
(\texttt{specmode = 'mfs'}) and the w-projection gridder with 768 $w$ planes.
Channel image frames within the three frequency ranges removed due to the presence of strong, 
persistent 
interference are replaced with NaNs.  The frequency ranges of contiguous good data are
887--993\,MHz, 960--1163\,MHz, 1299--1524\,MHz and 1630--1680\,MHz,
totalling 585\,MHz, or 74\% of the band.  

The MT-MFS broadband $Q,U$ images are provided for high sensitivity and 
higher angular resolution polarization images.  However a significant Faraday Depth will depolarize the broadband signals.  
The typical Faraday Depth of the polarized sources (Fig.~\ref{fig:FD_dist}) is much less than the FWHM of the RM
transfer function (Fig.~\ref{fig:rmtransfer}), so most sources are
detected in the MT-MFS images, if somewhat attenuated.  Reference
should be made to the observed Faraday Depth of a source when interpreting  the MT-MFS $Q,U$ values.

\begin{table*}
    \centering
    \begin{tabular}{|c|cc|rc|cccccc}
    \hline
          Pointing & \multicolumn{2}{c}{Coordinates} & \multicolumn{2}{c}{MFS Image} &  \multicolumn{5}{c}{RM Synthesis (smoothed cube)} \\
          &RA & DEC & \multicolumn{2}{c}{resolution} &  RMTF & median RM Synth RMS  & 
          \multicolumn{3}{c}{median per-chan bkgd RMS} \\
          &&&\multicolumn{2}{c}{(robust $=-0.50$)}& FWHM & off-source &Q& U&V \\
          &&&&& (rad\,m$^{-2}$) & ($\mu$Jy\,beam$^{-1}$)  & \multicolumn{3}{c}{($\mu$Jy\,beam$^{-1}$)} \\

    \hline
          XMMLSS\_12 &  02 17 51.0 & -04 59 59 & $7.75'' \times 6.81''$ & $-35.1^{\circ}$ & 56.8 & 3.12 & 25.3& 26.9& 25.1 \\
          XMMLSS\_13 & 02 20 42.0 & -04 49 59 &  $7.69'' \times 6.73''$ & $-21.2^{\circ}$& 56.8 & 3.97 & 27.9 & 31.2& 27.2 \\
          XMMLSS\_14 & 02 23 22.0& -04 49 59 & $7.93'' \times 6.90''$ & $-35.2^{\circ}$ & 57.6 & 2.90& 25.3 & 26.1& 25.4  \\
          COSMOS & 10 00 28.6&  +02 12 21 & $7.80'' \times 7.22''$ & $-20.1^{\circ}$ & 55.2 & 3.21 & 31.2 & 31.6& 31.5 \\
    \hline
    \end{tabular}
    \caption{Image and Cube Properties.}
    \label{tab:images}
\end{table*}

\section{Source Polarization}
\subsection{Spectro-polarimetry}

The official total intensity continuum images and catalogues for the
MIGHTEE early science fields are created by the MIGHTEE continuum working group.
These results have recently been published by \cite{Heywood_2022}.
For the purpose of measuring the polarization properties of the radio sources,
a total intensity source catalogue was generated for each pointing using pyBDSF on the Stokes $I$ plane of the broadband MFS images with robust = -0.5. 

Spectro-polarimetric data is derived for all sources within $0.5^{\circ}$ 
of the image centre (approximately the half-power point of the primary beam at band centre). 
For each radio source we extracted the $I,Q,U,V$ spectral profiles at the nearest pixel to the source position in the smoothed cube.  The 18$''$ synthesized beam is heavily over-sampled (12 pixels per FWHM), so the amplitude of the nearest pixel to the source position accurately represents the
peak intensities.

The image cubes contain low-level, diffuse polarized signals arising from 
differential Faraday rotation of the Galactic synchrotron background.  
To remove this effects from the compact source spectra,  
background spectra ($Q_{\rm bkgd}$, $U_{\rm bkgd}$) were derived for each source from the average $U$ and $Q$ signal in an annular box around the source in each frequency channel. The width of the annulus is 26 
pixels (39$''$).   The half-width of central hole is 
$\Delta = 4\,\theta_s$,
where $\theta_s$ is the semi-major axis of the source from the pyBDSF fit in the MFS image.
The area of the annular box is $\sim$39 beam solid angles.  The standard deviation of
the values in the background box was used as an estimate of the local rms per channel around the source.  The median per-channel background RMS before over all sources and channels
before primary beam correction is given for $Q$, $U$ and $V$ in Table~\ref{tab:images}. The per-channel RMS is
typically 25-30 $\mu$Jy\,beam$^{-1}$, and the $Q$ and $U$ values agree to within 4\%, except
for XMMLSS\_13 where the difference is $\sim$10\%.  XMMLSS\_13 is more affected by a 
strong, off-axis source that appears to affect the background RMS in $U$ slightly more than $Q$.

\subsection{Polarization Leakage}
\label{sec:leakage}
\begin{figure*}
    \centering
    \includegraphics[width=\textwidth]{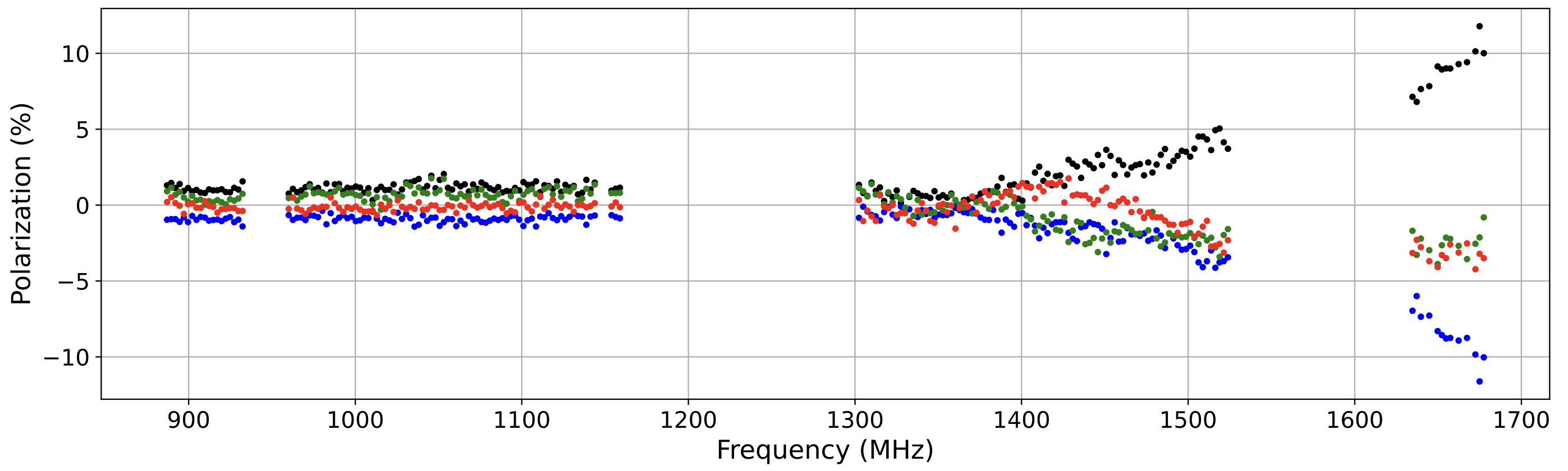}
    \caption{Percent polarization in $Q$, $U$, $V$ (blue, green, red) and
    total linear polarization (black) for a strong
    source in XMMLSS\_13 at a distance of 0.45$^{\circ}$ from the field centre.
    The high circular and linear polarization leakage above about 1400\,MHz for this off-axis sources arises
    from the large differential $X-Y$ beam squint above this frequency, \citep{asad_2021,Sekhar2022}.}
    \label{fig:off_leakage}
\end{figure*}

On-axis leakage residuals after calibration are measured to be about 0.1\%.  
However, there is frequency-dependent beam squint and
squash at the high-end of the L-band
\citep{asad_2021,Sekhar2022}, that leads to off-axis frequency-dependent leakage.
The differential squint between the beams creates instrumental polarization that
grows to levels of several percent at the high end of the band. 
This effect is demonstrated in Figure~\ref{fig:off_leakage} which shows
the percent polarization in $Q, U$ and $V$ for a strong source (13 mJy) at a distance
of 0.45$^{\circ}$ from the field centre in XMMLSS\_13.  
Above about 1400\,MHz  the polarization grows quickly with frequency to several percent in all polarizations.   
To avoid the strong leakage at the upper end of the band we 
restrict the analysis the polarization signals to 
frequencies below 1380\,MHz.  
Below this frequency the leakage remains less than 0.2\% within 
0.5$^{\circ}$ of field centre.

\subsection{Detection of Polarized Emission}
\label{sec:pol_detections}
To search for significant polarized emission from the radio sources, Faraday depth spectra were constructed from the background-corrected Stokes spectra of each total intensity source,  
as well as for one nearby off-source position that is offset 
from the source by $\Delta$.
The source density in the total intensity image is quite high.  
Care was taken to ensure that the off-source position did not 
coincide with a source that has total flux density greater than 100$\mu$Jy\,bm$^{-1}$.
The Faraday depth spectrum is calculated as
\begin{equation}
    F(\phi) = \biggl(\frac{1}{\sum_{j=1}^N w_j}\biggr ) \sum_{j=1}^N P_j e^{-2 i \phi  (\lambda_j^2-\overline\lambda^2)} \cdot w_j.
    \label{eqn:faraday}
\end{equation}
Here the sum is over all channels, $j$, $\overline\lambda$ is the mean wavelength and
\begin{equation}
    P_j = (Q-Q_{\rm bkgd})_j - i (U-U_{\rm bkgd})_j.
    \label{eqn:background}
\end{equation}
The RM synthesis weights are 
\begin{equation} \label{eqn:weights}
    w_j = \biggl (\frac{1}{\sigma_{QU_j}} \biggr )^2,
\end{equation}
where $\sigma_{QU_j}$ is the average of the $Q$ and $U$ local background RMS in channel $j$.

\begin{figure}
    \centering
    \includegraphics[width=\columnwidth]{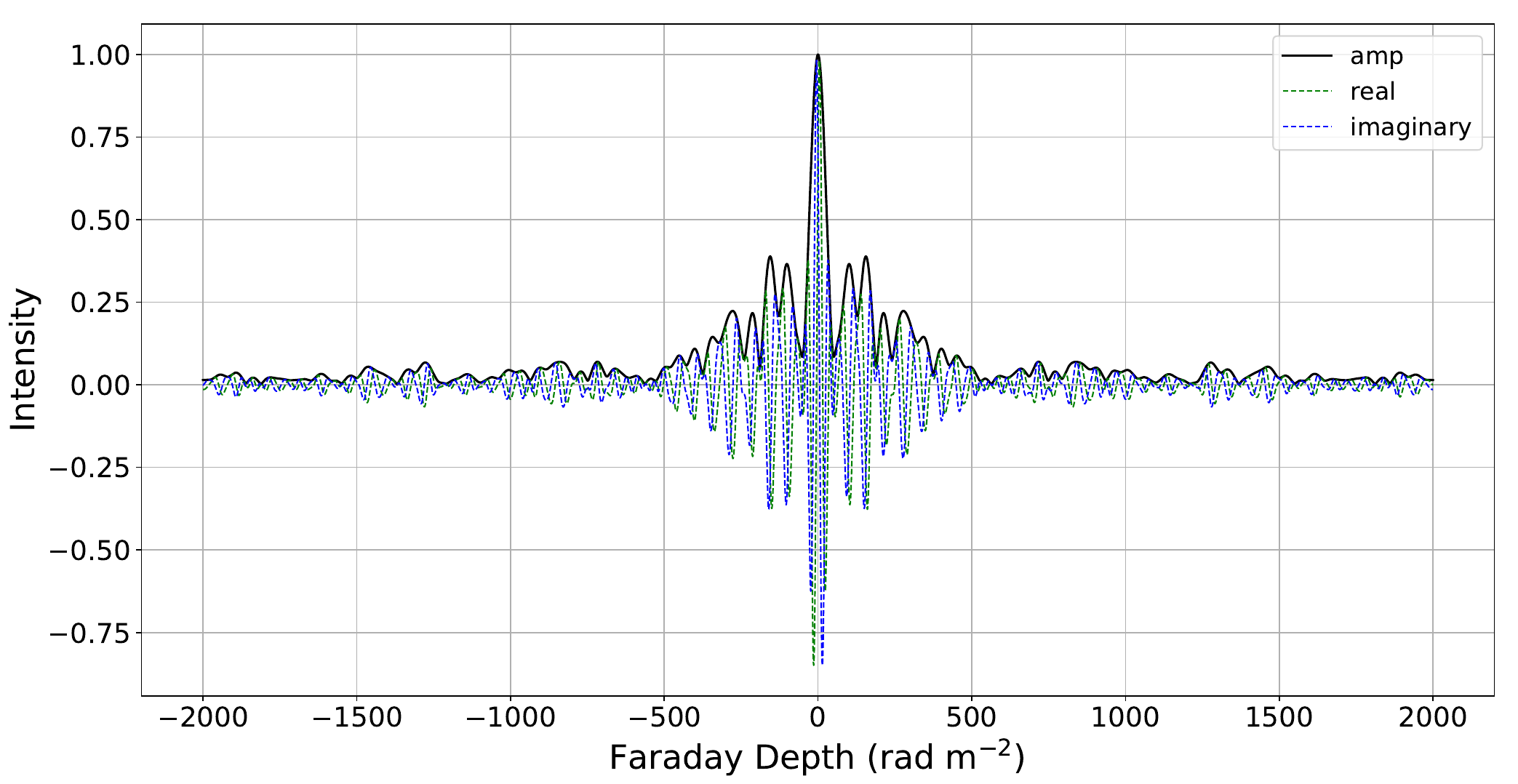}
    \caption{Typical RM transfer function for frequencies from 887-1380\,MHz.
    The solid black line is the amplitude. The missing sections of the band result in high sidelobes of about 40\%.}
    \label{fig:rmtransfer}
\end{figure}

Faraday depth spectra were constructed, using data at frequencies less than 1380\,MHz, over a Faraday Depth range of $\pm$2000\,rad\,m$^{-2}$.  
Figure~\ref{fig:rmtransfer} shows a characteristic RM transfer function.
The missing frequency range from 1163 --1299\,Mhz results in 
strong sidelobes of about 40\% at an offset of about 
$\pm$90\,rad\,m$^{-2}$ from the main lobe.  
The main lobe of the transfer function has a typical width of 56\,rad\,m$^{-2}$.  Precise values of
the FWHM of the RM transfer function are given for each
field in Table~\ref{tab:images}.
With a channel width of 2.51\,MHz, the maximum detectable rotation measure is
about 5,000\,rad\,m$^{-2}$ at band centre. For a maximum frequency of 1380\,MHz, the
largest detectable width for a contiguous Faraday depth feature is 67\,rad\,m$^{-2}$.
The noise in the RM synthesis spectrum was estimated from
standard deviation of the real part of the off-source
spectrum.  Table~\ref{tab:images} lists the median
off source RMS over all sources for each field.  
The typical value is $\sim$3\,$\mu$Jy\,bm$^{-1}$, consistent
with the RMS of the total intensity images.

\begin{figure*}
    \centering
    \includegraphics[width=\textwidth]{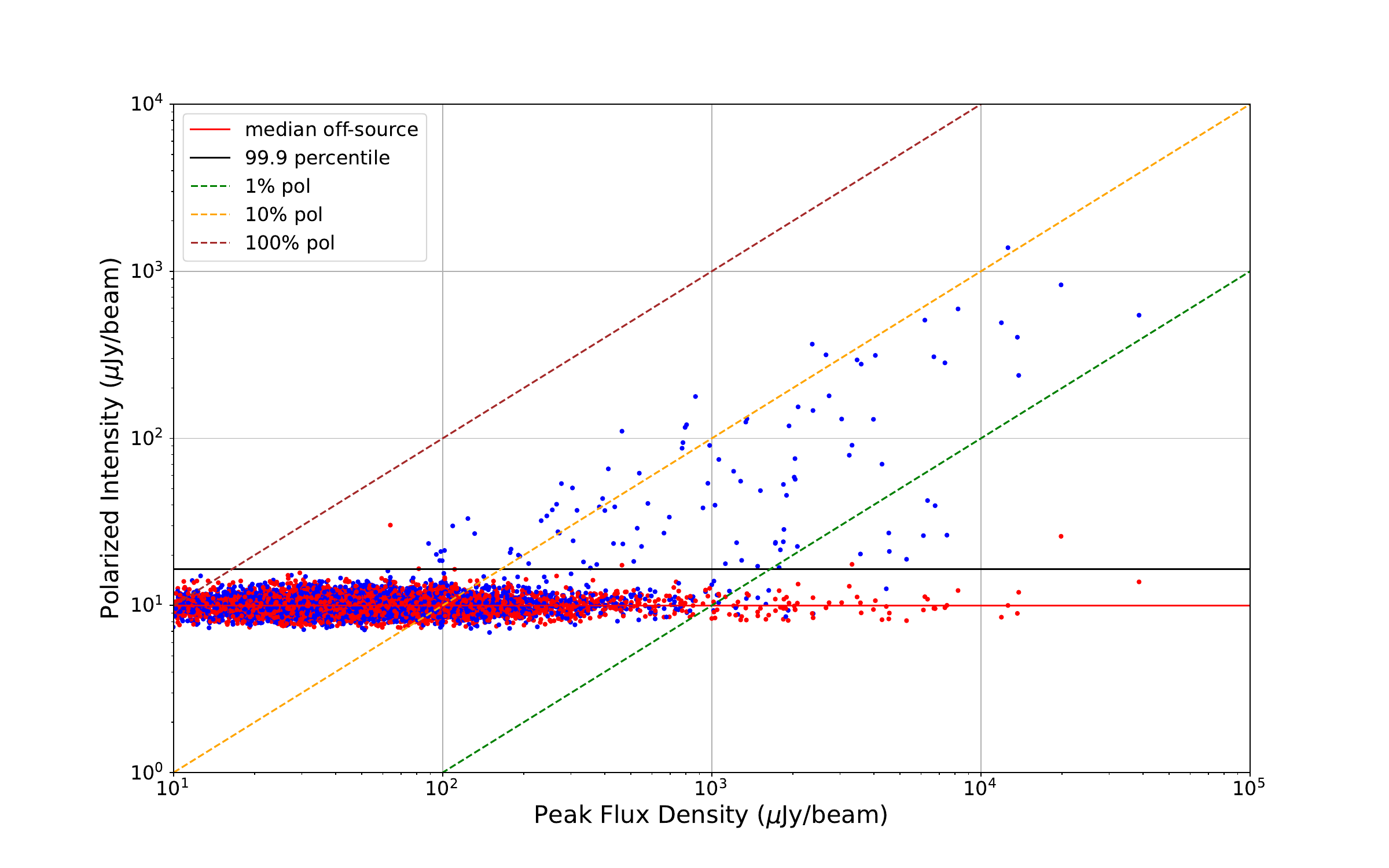}
    \caption{Polarized intensity of the maximum of the RM synthesis spectra for the on-source 
    (blue dots) and off-source (red dots) positions plotted against the peak total intensity 
    for each source. This plot is for the COSMOS field.  
    The diagonal dashed lines represent
    lines of constant fractional polarization of 1, 10 and 100\%.  The horizontal red line
    is the median value of the off-source spectral peaks. The horizontal black line shows the 99.9 percentile of the
    off-source spectral peaks at 16.5\,$\mu$Jy\,bm$^{-1}$.}
    \label{fig:on-off_COSMOS}
\end{figure*}

\begin{figure*}
    \centering
    \includegraphics[width=\textwidth]{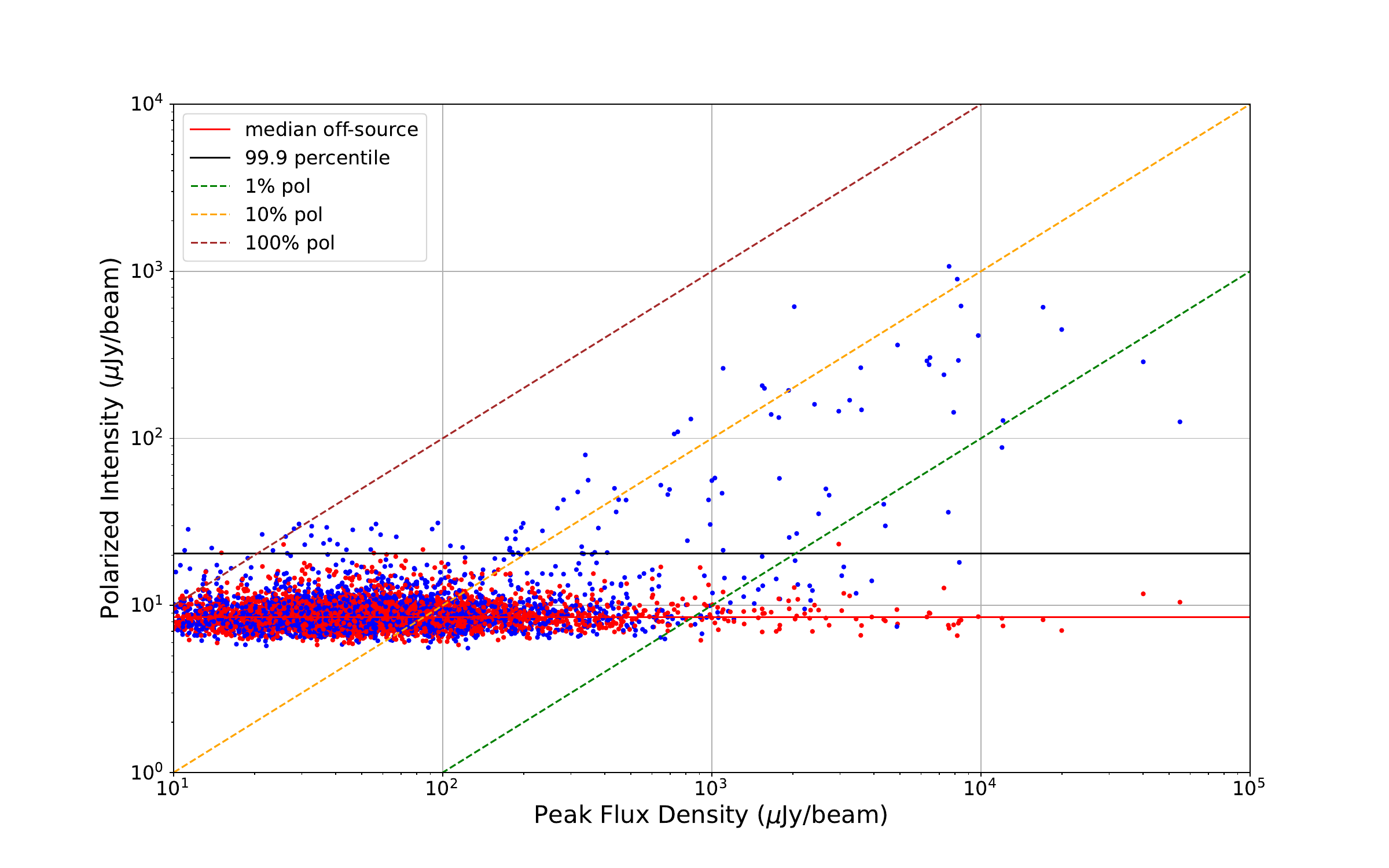}
    \caption{As in Figure~\ref{fig:on-off_COSMOS}, but for XMMLSS\_14. Note the clustering
    of blue dots are very faint peak total intensity with very high percentage polarisation
    that are not seen in the COSMOS data.
    These points arise due to the presence of faint, structured diffuse emission from the Galactic foreground in the
    XMMLSS fields.}
    \label{fig:on-off_XMMLSS_14}
\end{figure*}
The polarized intensity and Faraday depth of the strongest
component in both the on-source and off-source spectra was
measured for each source.
The distribution of the amplitudes of the off-source spectra provides a statistical measure of the probability
of detecting an RM synthesis peak of a given amplitude from the noise of the Stokes $Q$ and $U$ planes of the image cubes. 
Figures~\ref{fig:on-off_COSMOS} and \ref{fig:on-off_XMMLSS_14} show the distribution of the peak amplitude of the RM synthesis spectra
for the on- and off-source positions plotted against the peak total intensity for each source.
The black horizontal line in each Figure is the 99.9 percentile of the off-source spectral
peaks. On-source peaks (blue dots) that fall above this line are taken to be detected polarization signals.
Table~\ref{tab:detections} lists the total number of total
intensity sources with 0.5$^{\circ}$ of field centre, the polarization detection thresholds and the raw number of detected polarization signals
for each of MIGHTEE early science fields.

The three fields observed in
XMMLSS have significant overlap.  We thus created a merged catalogue in 
which duplicate detections are removed.  
For objects detected in more than one field, we retained only the
instance from the field in which the sources is closest to the field centre.
The merged XMMLSS catalogue has 9431 total intensity radio components.
Despite the local background removal (Eqn~\ref{eqn:background}), 
the XMMLSS results remain affected at low flux density by residual faint diffuse 
emission from the Galactic foreground.  This results in spurious detections of polarized emission at low levels.
The effect can be seen in Figure~\ref{fig:on-off_XMMLSS_14} as blue
dots that rise above the detection threshold at peak flux densities
below $\sim$100\,$\mu$Jy\,bm$^{-1}$.  These sources have very high
fractional polarization.  To remove them we have applied a filter
in total flux density and fractional polarization.
This results in 216 polarization detections in the merged XMMLSS 
source list.
\begin{table}
    \centering
    \begin{tabular}{|c|c|c|c|}
    \toprule
          Field &  Total Sources & Detection Threshold & Detected \\
          & (r < 0.5$^{\circ}$) & ($\mu$Jy\,bm$^{-1}$) & \\
    \hline
          XMMLSS\_12 & 3695 & 27.4 & 121 \\
          XMMLSS\_13 & 3368 & 31.0 & \ \ 89  \\
          XMMLSS\_14 & 3405 & 27.4 &  143   \\
          COSMOS &  3840& 16.5 & 108 \\
    \bottomrule
    \end{tabular}
    \caption{Polarized Intensity detection thresholds and number
    of detected polarized sources in each MIGHTEE early science field}
    \label{tab:detections}
\end{table}

\subsection{Polarization and Faraday Depth Measurement}
\label{sec:polarization}
Following the analysis to measure detections of polarized emission. 
The Faraday synthesis was redone for each source in fractional 
polarization.
For this purpose the observed (before primary beam correction) Stokes $I$ peak flux
spectrum over the full 880 to 1680\,MHz band is fit by a power law function of the form
\begin{equation}
    I(\nu) = I_o \biggl(\frac{\nu}{\nu_o}\biggr )^{\alpha + C \ln{(\nu/\nu_o)}}.
    \label{eqn:specfit}
\end{equation}
The quantity $P_j$ in equation~\ref{eqn:faraday} is replaced by
\begin{equation}
    p_j = \frac{(Q-Q_{\rm bkgd})_j - i (U-U_{\rm bkgd})_j}{I(\nu_j)}.
    \label{eqn:pol}
\end{equation}
and the weights become
\begin{equation} \label{eqn:QU_weights}
    w_j = \biggl (\frac{I(\nu_j)}{\sigma_{QU_j}} \biggr )^2.
\end{equation}
 
For each source the amplitude (percent) and the position of the
dominant component of the Faraday depth spectra was recorded.
The effective frequency of the measurements is taken
as the weighted mean frequency of the channels
used.
The uncertainty on Faraday depth can be calculated as 
\begin{equation}
    \Delta_{\phi} = \frac{\delta \phi}{2|P_{\rm peak}|/\sigma_{\rm snyth}},
\end{equation}
where $\delta \phi$ is FWHM of the RM transfer function, $P_{\rm peak}$ is the peak amplitude and $\sigma_{\rm synth}$ is the noise in the
RM synthesis spectrum \citep{bb05}, measured as the RMS of the real component of the off-source spectrum.

\begin{figure}
    \centering
    \includegraphics[width=0.45\textwidth]{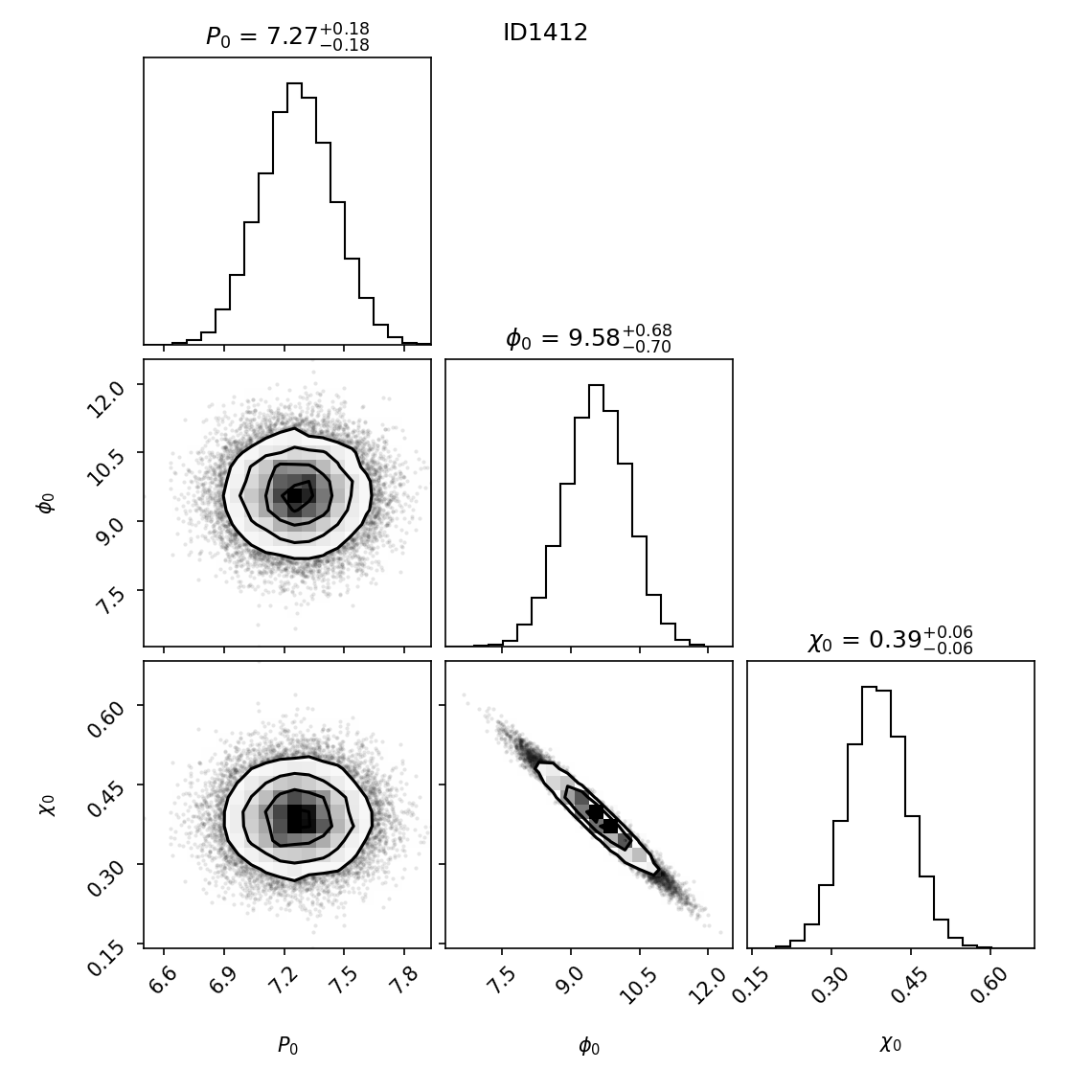}
    \caption{Posterior probability distribution from QU-fitting of source ID\,1412 in the XMMLSS field. Parameters are: $P_0$, the polarization percentage; $\phi_0$, the Faraday depth; and $\chi_0$, the intrinsic polarization angle. Contours show confidence intervals from 1 to 3\,$\sigma$.}
    \label{fig:qufit}
\end{figure}

For polarization detections we also derived Faraday depth values for 
the strongest component in each spectrum derived using the QU-fitting technique on the background-subtracted QU data\footnote{\url{https://github.com/as595/QUFitting}}. The recovered rotation measure for each source is represented by a maximum-a-posteriori (MAP) expectation value obtained using MCMC optimisation of a single Faraday-thin component model. In each case the MAP value was obtained using an MCMC with a randomly initialised burnin of 5,000 samples per chain, followed by a production run of 10,000 samples initialised from the maximum-likelihood (ML) value of the burnin. The MCMC uses uniform priors in the range $0<P_0<100\,\%$, $-1,000<\phi_0<+1,000$\,rad\,m$^{-2}$ and $-\pi/2\leq \chi_0 < \pi/2$\,radians. Following the burnin, the $\chi_0$ prior is relocalised to cover $\pi$\,radians around the ML initialisation point. MCMC convergence was assessed using the auto-correlation length of the chains; a burn-in of 5 times the autocorrelation length was discarded and chains were thinned by a factor of 15 in order to calculate final posterior distributions. Posterior uncertainties at a level of 1-$\sigma$ are provided for MAP estimates. An example of the posterior distributions obtained from QU-fitting is shown in Figure~\ref{fig:qufit} for the XMMLSS source ID\,6607. The relationship between the rotation measure estimates recovered from QU-fitting and those measured directly from the Faraday depth spectra is illustrated in Appendix~\ref{app:qufit}. 

\begin{figure}
    \centering
    \includegraphics[width=0.95\columnwidth]{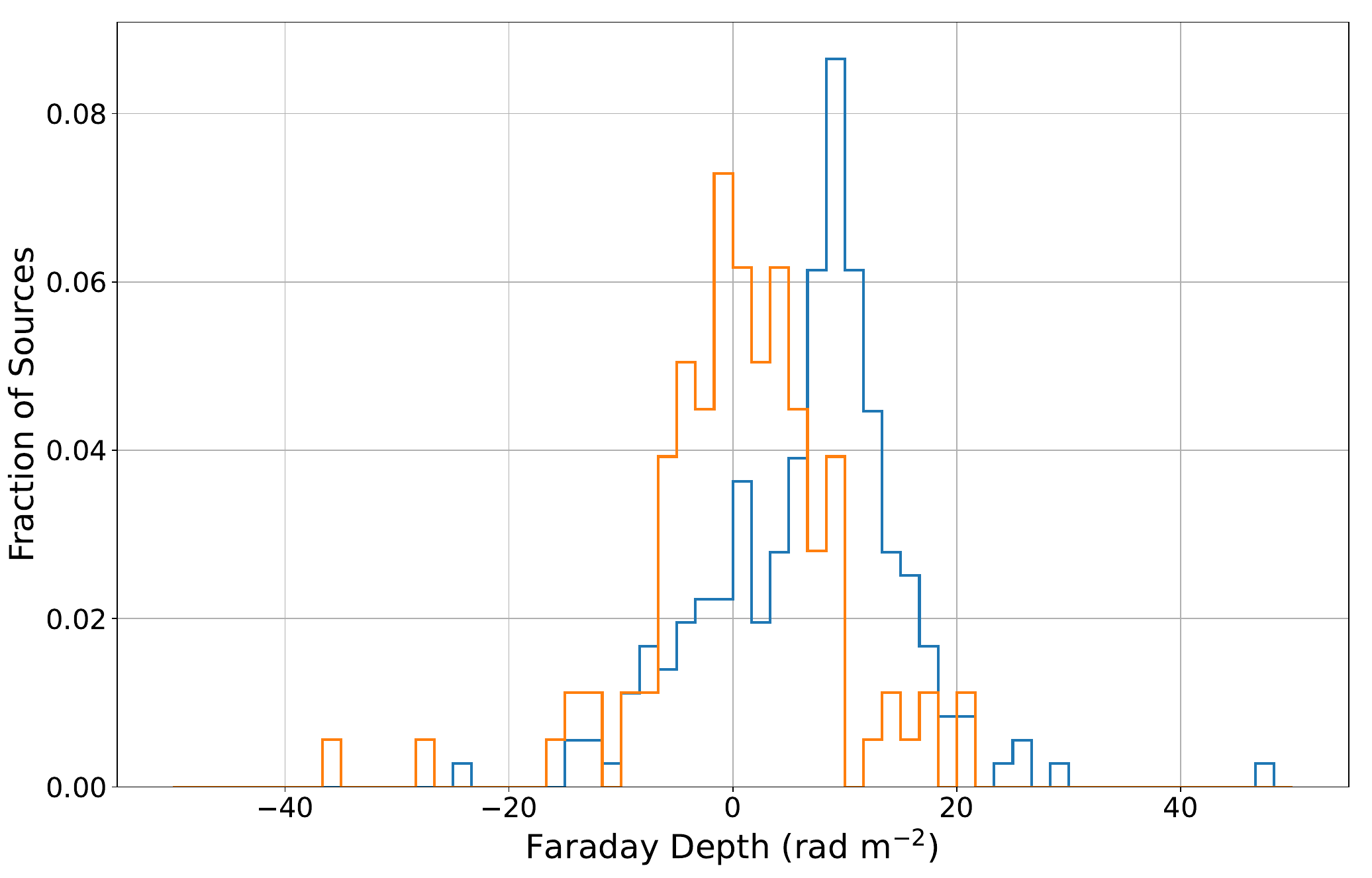}
    \caption{Distribution of Faraday Depth of the dominant Faraday Depth component for polarized sources in  XMMLSS (blue) and COSMOS (orange). 
    }
    \label{fig:FD_dist}
\end{figure}

Figure~\ref{fig:FD_dist} shows the distribution of Faraday Depth of
the dominant Faraday depth spectral component for each of the polarized sources for
the COSMOS and XMMLSS data.  The median value of the COSMOS 
*Faraday Depth is 1.03 rad\,m$^{-2}$, with a RMS estimated from the 
Median Asbolute Deviation of 6.60 rad\,m$^{-2}$.
For XMM-LSS the median is signficantly higher at 8.24 rad\,m$^{-2}$
but with an essentially identical RMS of 6.56 rad\,m$^{-2}$.
The enhanced median Faraday depth of XMM-LSS is due to the Galactic foreground Faraday screen.  

\citet{hutschenreuter2022} created an interpolated all-sky image of 
the foreground Galactic Rotation Measure (GRM) based on a compilation 
of data sources with an average density of about 1 RM per
square degree.  The median values of the MIGHTEE Faraday depths are in general agreement with the mean values over each field from
\citet{hutschenreuter2022} of $-0.5\pm7.6$ for COSMOS and
$6.7\pm$ 5.7 for XMM-LSS.

More local estimates of GRM can also be made for the area around each MIGHTEE source, $i$, using the polarised sources that are within a radius, $R$, from that source. We take the median of all sources within that radius to estimate the local median RM, $\Phi_{i,R}$. The local uncertainty for each estimate, $\Delta_{\Phi_{i,R}}$, is calculated from the median absolute deviation of the selected sources.

We define a quantity $\eta$:
\begin{align}
    \eta &= \frac{1}{N}\sum^N_{i=1}\begin{cases}
              1 &\text{if} \,\, \left| \Phi_{i,R} - \Phi_{i,\text{H+}}\right| < \sqrt{\Delta_{\Phi_{i,R}}^2 + \Delta_{\Phi_{i,\text{H+}}}^2}  \\
              0 &\text{else.}\\
         \end{cases}, \label{eqn:grm_alignment}
\end{align}
where $\Phi_{i,\text{H+}}$ and $\Delta_{\Phi_{i,\text{H+}}}$ are the value of the GRM and GRM error at the position of source $i$ from \citet{hutschenreuter2022}, and $N$ is the number of sources in a given field. Conceptually, $\eta$ is the fraction of local GRM values that are within 1-$\sigma$ of the current literature GRM values. 

\begin{figure}
    \centering
    \includegraphics[width=0.95\columnwidth]{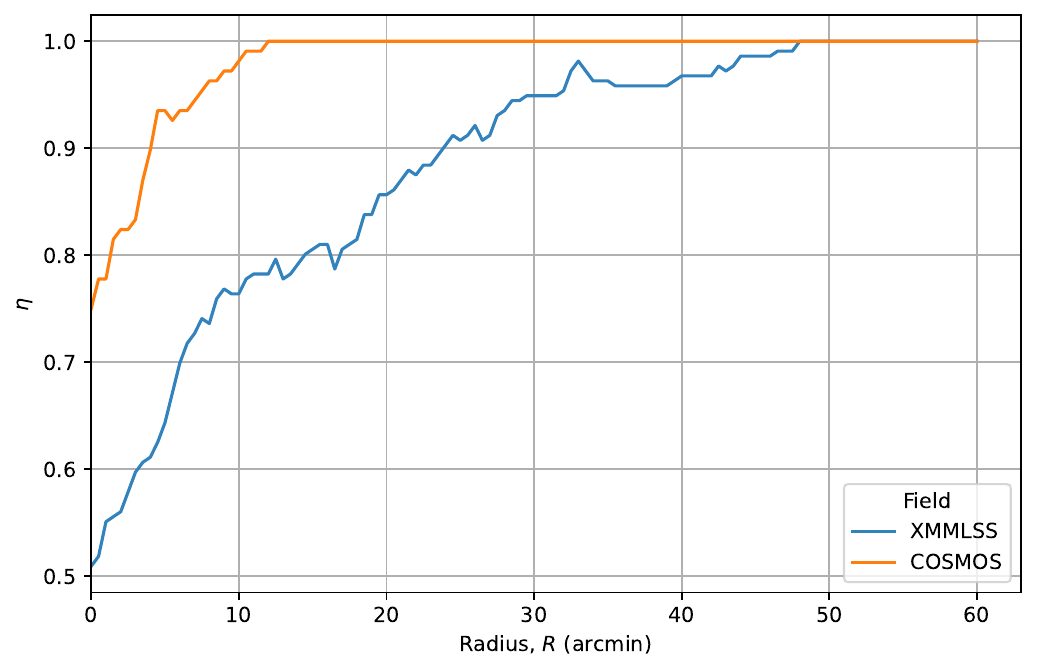}
    \caption{The fraction, $\eta$, of MIGHTEE-POL Galactic RM estimates which lie within 1$\sigma$ of the GRM  value from \citet{hutschenreuter2022}, as a function of the radius, $R$, used to estimate the local GRM.
    }
    \label{fig:grm}
\end{figure}

We use $\eta$ as a function of the radius, $R$, with which we select sources to include in the median value to highlight the convergence of the local GRM estimates to the lower resolution GRM estimates from \citet{hutschenreuter2022} as $R$ increases. This is illustrated in Figure~\ref{fig:grm}, which shows a clear convergence of both fields to the literature values as $R$ approaches 1 degree.
Deviations from the literature value for smaller values of $R$ imply
significant angular structure of the foreground RM on sub-degree scales.  This effect is stronger in XMM-LSS where the average foreground is larger compared to COSMOS. 
Studies that rely on data sampled on degree scales to
remove foreground RM estimates will suffer significant uncertainty. 
A more detailed analysis these previously unseen Galactic RM structures
will be presented in future work.

\section{Spectral Index}
Equation~\ref{eqn:specfit} was used to fit the total intensity spectrum for all sources to derive an internal spectral index. 
Before fitting the peak total intensities in each channel are corrected for the primary beam variation in direction and frequency using the katbeam python module\footnote{\url{https://github.com/ska-sa/katbeam}}, which uses the circularly symmetric cosine-squared formulation described in \citet{Mauch_2020}.

\begin{figure}
    \centering
    \includegraphics[width=1\columnwidth]{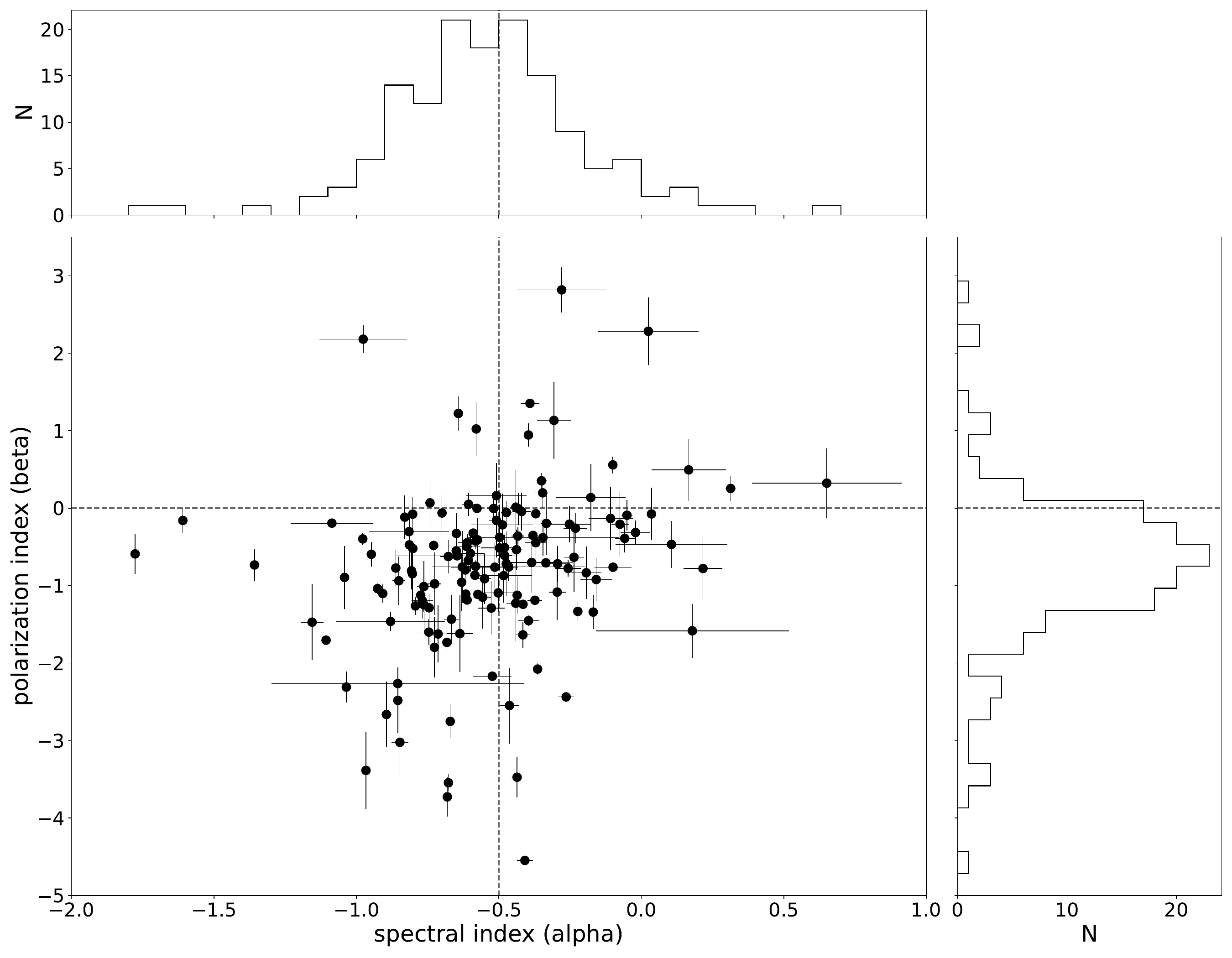}
    \caption{Fractional polarization spectral index, $\beta$, versus total
    intensity spectral index, $\alpha$, for 143 polarized sources with errors $\sigma_{\alpha} < 0.3$ and $\sigma_{\beta} < 0.5$.  The top and side panels show
    the distributions in $\alpha$ and $\beta$.
    The vertical and horizontal dashed lines divide the
    plot into quadrants at $\alpha = -0.5$ and $\beta = 0$.
    }
    \label{fig:alpha_beta}
\end{figure}

For polarization detections a polarization spectral index, $\beta$ 
is calculated by fitting a power law $p(\lambda) \propto \lambda^{\beta}$ to the percent polarization $p(\lambda)$.
Before fitting, following \citet{George_2012},
the $p(\lambda$) from Equation~\ref{eqn:pol} for each channel is
bias corrected using $\sigma_{QU}$,
\begin{equation}
p(\lambda_i)_{\rm bias} = \sqrt{p(\lambda_i)^2 - 2.3\sigma_{QU_i}^2}.
\end{equation}
Sources with $\beta$<0 are "depolarized" with lower
fractional polarization at lower frequencies.  This
behaviour is generally attributed to disordered magnetic fields or to internal or external Faraday depolarization.  A overview
of spectral models of depolarization processes can be found in \cite{Farnes_2014},
who compiled and modelled the polarized spectral energy distributions derived from 
the literature for 951 strong sources.  The polarized flux limit of the \cite{Farnes_2014} is based on the NVSS RM catalogue \citep{Taylor_2009},
about 3\,mJy at 1400\,MHz.  Sources with $\beta > 0$ are so-called ''repolarized''
sources, with fractional polarization increasing at longer wavelengths.

Due to the limited frequency range (880-1380\,MHz) for the Faraday synthesis,
the derivation of $\beta$ has high uncertainty for the faintest polarized sources. In Figure~\ref{fig:alpha_beta} we plot $\beta$ versus the total intensity spectral index $\alpha$ for 143 polarized sources for which the 1$\sigma$ error on 
$\beta$ is less than 0.5 and the error on $\alpha$ is less than 0.3. 
The Figure shows a similar result to the findings of \cite{Farnes_2014} 
(see their Figure 7).  There is a trend of stronger depolarization toward
sources with more negative spectral index.  Repolarization is observed for 13\% of the sources, and predominantly for objects with flatter spectral 
index.
The relationship between $\alpha$ and $\beta$ emphasizes the conclusion that
depolarization and repolarization are intrinsic properties of the emitting sources, 
as opposed to external effects.

\section{Redshifts}
The radio sources were cross-matched with photometric and spectroscopic redshift data from \citet{Adams2021} and \citet{hatfield_hybrid_2022}.
Additional spectroscopic redshifts from \citet{Vaccari2022} were cross-matched.  

Each redshift catalogue was cross-matched with the radio source catalogue  utilising the python astropy function \texttt{match\_coordinates\_sky}.
The catalogues are cross-matched within a matching radius $d_\mr{match}$
depending on radio source position uncertainties $\sigma_{\rm ra}$ and
$\sigma_{\rm dec}$ via $d_{\mr{match}} = x_\sigma \cdot d_\sigma$  where
\begin{equation}
  d_\sigma = \sqrt{\sigma_{\rm ra}^2+\sigma_{\rm dec}^2 + 2\sigma_{\rm opt}^2 } \quad \:.
\end{equation}
The source positions uncertainties of the redshift catalogues, which based on optical observations, are assumed to be $\sigma_{\rm opt}=0.3"$.
The scaling factor $x_\sigma$ was calculated to maximize reliability 
and completeness by performing a statistical
matching analysis similar to the approach of \cite{Farnes_2014} and \cite{Mauch2007}.
The actual match count is compared to the match count between the 
same catalogues with randomised source positions as a function of $x_\sigma$ to determine the largest matching radius at which match counts still differ from match counts with randomised positions.
At that radius we estimate the completeness by counting the first neighbour matches as well as the fraction of spurious source associations by counting the matches of the second neighbour matches.
The \citet{Adams2021} catalogue has a high source density.
We used $x_{\sigma} = 1.5$, resulting in matches for 81.1\,\% (3.3\,\% spurious) of the radio sources in XMM-LSS and 81.4\,\% (5.3\,\% spurious) in COSMOS.  For \citet{Vaccari2022} a value of $x_\sigma$ = 4.5 provided matches for 40.6\,\% (6.0\,\% spurious) of the radio sources in XMM-LSS and 56.4\,\% (10.1\,\% spurious) in COSMOS.

\begin{figure}
    \centering
    \includegraphics[width=1\columnwidth]{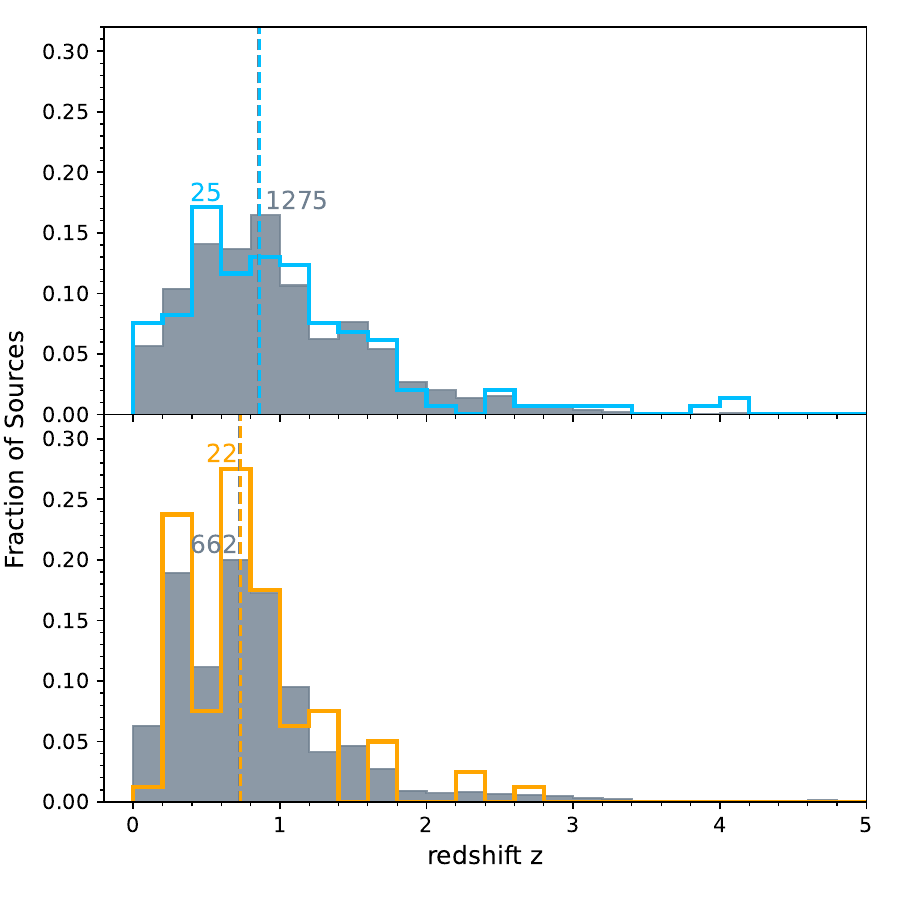}
    \caption{Redshift distribution for XMM-LSS (top) and COSMOS (bottom).
    The plot shows the fraction of sources against the best redshift estimate.
    For XMM-LSS 7753 radio sources have corresponding redshifts (top
    gray histogram) of which 216 are detected in polarised emission (blue).
    In the COSMOS field 3308 radio sources have corresponding redshifts (bottom
    gray histogram) of which 81 show polarised emission (orange).
    The dashed lines indicate the median redshift for each distribution; it is virtually identical between the polarised sources and the total population in each field. The number on top of the highest bin shows the source count in that bin.
    }
    \label{fig:redshifts}
\end{figure}

Figure~\ref{fig:redshifts} shows the redshift distribution for all redshift-associated sources in XMM-LSS and COSMOS. 
Following \citet{Duncan2022} we exclude photometric redshifts $z_\mr{phot}$ with very high uncertainties
\begin{equation}
\label{eq:duncan}
  \sigma_z/(1+z_\mr{phot}) > 0.2 \:.
\end{equation}
This affects 849 sources in XMM-LSS and 279 in COSMOS.
Of the 324 detected polarised radio sources, 227 have associated redshifts, see Table~\ref{tab:redshifts} for details.
The redshift distribution of the polarized detections are shown as the solid
lines in Figure~\ref{fig:redshifts}.
The polarized sources are seen at all redshifts and follow the same redshift distribution as the total radio source population.
The median redshifts for both populations in XMM-LSS and COSMOS are 
virtually identical. 
XMM-LSS: $z_\mr{med}=0.8576$,
$z_\mr{med,pol}=0.8585$; COSMOS:
$z_\mr{med}=0.7280$,
$z_\mr{med,pol}=0.7292$.

\begin{table}
    \centering
    \begin{tabular}{|l|c|c|c|}
    \toprule
          Field & Total & Spectroscopic z & Photometric z \\
    \hline
          XMM-LSS & 146 & \ \ 92 &  54 \\
          COSMOS & \ \ 81 & \ \ 69 &  12 \\
    \hline
          Total & 227 & 161 &  66 \\
    \bottomrule
    \end{tabular}
    \caption{Number of polarised sources with spectroscopic and
    photometric redshift for XMM-LSS and COSMOS.}
    \label{tab:redshifts}
\end{table}

\begin{figure}
    \centering
    \includegraphics[width=1.0\columnwidth]{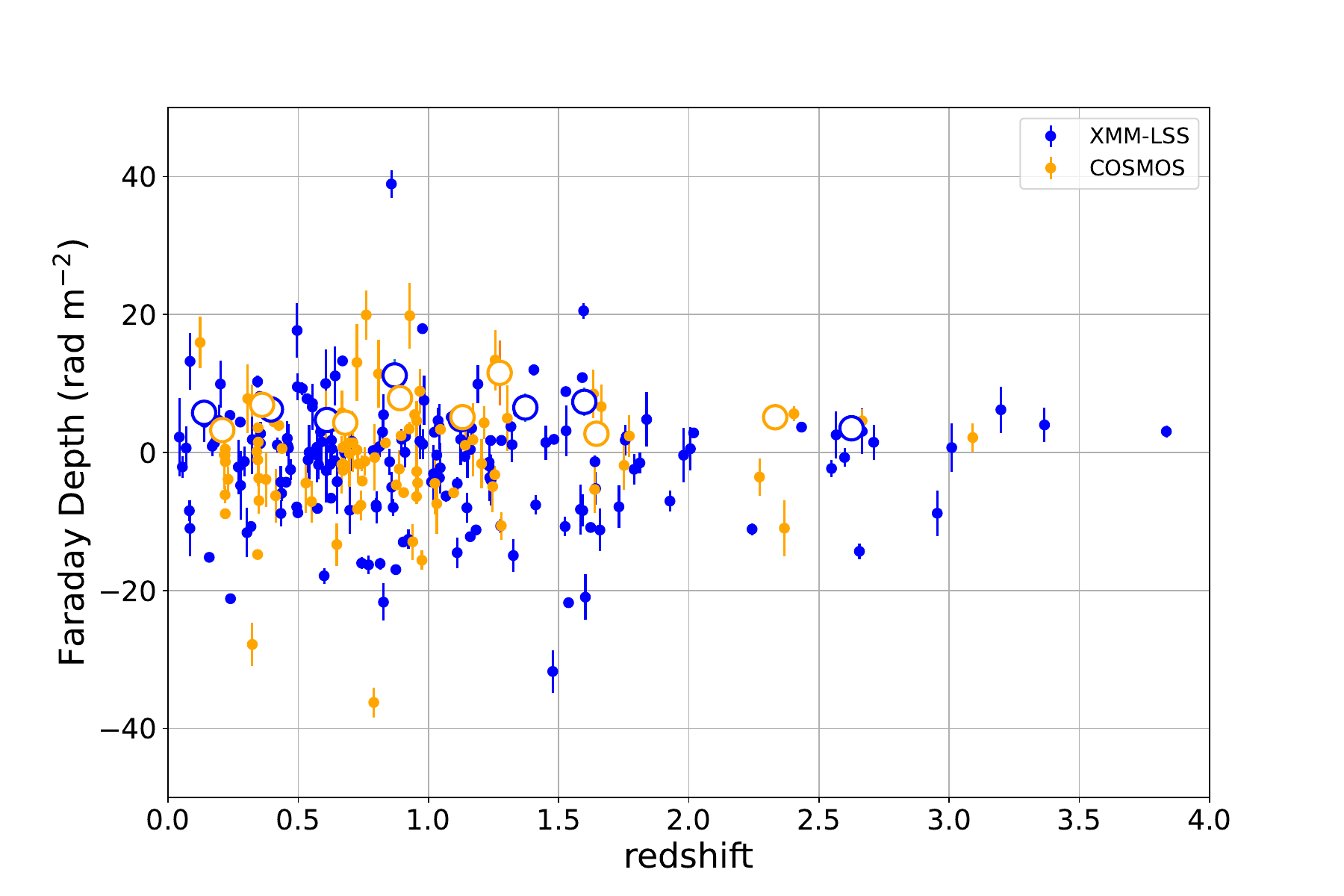}
    \caption{Faraday depth of the dominant Faraday component, with Galactic foreground estimate removed, against redshift for XMM-LSS (blue) and COSMOS (orange). 
    Large open circles are the standard deviation of the
    Faraday depths in redshift bins. Bin widths are
    $\Delta z$ = 0.25 out to $z=1.75$.  The final bin contains all sources with $z > 1.75$.
    }
    \label{fig:faraday_depth_vs_redshift}
\end{figure}

To reduce the effect of the foreground GRM on the 
Faraday depths, we define an average Faraday depth 
for all sources within a given pointing, and subtract
the average from each source in that pointing.
Figure~\ref{fig:faraday_depth_vs_redshift} shows the 
foreground-removed Faraday depth of the dominant Faraday synthesis component for each polarized sources as
a function of redshift. 
Because of the wide bandwidth and high sensitivity of the MIGHTEE observations, the error on the measurement of Faraday depth is quite low, from less than 1 to a few rad\,m$^{-2}$.
The scatter in the redshifts is thus due to a combination of
scatter in the intrinsic Faraday depth of the radio sources, or 
small scale spatial variation in the Galactic foreground screen.
The large open circles in Figure~\ref{fig:faraday_depth_vs_redshift}
show the standard deviation of the Faraday depths in redshift
bins. The bin width is $\Delta z = 0.25$ out to $z = 1.75$. The final bin contains all sources with $z > 1.75$. 
To reduce the impact of outliers, the standard deviations are estimated from the Median Absolute Deviation.
There is no evidence of a trend in the magnitude of the 
standard deviation of the Faraday Depths with redshift. 
The average standard deviation is $5.9\pm2.7$ for COSMOS and
$6.3\pm2.2$ for XMM-LSS.  The slightly larger value for XMM-LSS may reflect an increased contribution from scatter in the GRM, as suggested in Figure~\ref{fig:grm}.

\cite{hammond_2013} catalogued redshifts for 4003 sources from the 
\cite{Taylor_2009} NVSS 1.4\,GHz RM catalogue.  
While the typical error on
the NVSS RMs is much larger (10 to 20 rad\,m$^{-2}$), \cite{hammond_2013} also
found no evidence for redshift evolution in RMs corrected for the Galactic foreground.  
Using very precise RM from the LOFAR Two-metre Sky Survey Data Release 2 \citep{Shimwell_2022}, \cite{Pomakov_2022} measured an average
difference in $|RM|$ of $1.79\pm0.09$ between pairs of objects at different redshifts that share the same line of sight to within 30 arcminutes. This difference is independent of the redshift difference between the
pairs.
The structure in Galactic RM seen in our data
imply that at least some or all of this signal may arise from differential GRM on small scales.

\cite{Bernet_2012} similarly found no evidence for an increase in the dispersion of Faraday depths with redshift using the \cite{Taylor_2009} data.
This is despite detecting an increase of dispersion from Faraday depths derived at 5 GHz.
To account for this difference, \cite{Bernet_2012} model the observed RM as the sum of the foreground GRM and a contribution from an inhomogeneous Faraday screen at the source.  The polarized emission from the source is described as 
\begin{equation}
p = p_o \bigl (f_c \exp(-2\sigma_{\rm RM}^2(1+z)^{-4}\lambda^4) + (1-f_c) \bigr ).
\end{equation}
Here $f_c$ is the covering factor of the screen over the source,
$\sigma_{\rm RM}$ is the Faraday dispersion of the screen and
$\lambda$ is the radiation wavelength.
As shown by \cite{Bernet_2012} for low frequencies the portion
$f_c$ of the polarized flux at source that propagates through the Faraday screen is depolarized.  
We have carried out some Monte Carlo simulations of this model as a function of redshift and find that the partial Faraday screen at source 
introduces additional scatter at low redshifts.  The scatter
is reduced at high redshift due to the shorter wavelength of the
radiation at source.  This effect runs counter to the expected
increase with redshift in the magnitude of RM and may obscure 
its detection with low-frequency observations.

\section{Catalogues and Data}

Catalogues of spectra-polarimetric data and Faraday synthesis result are provided for each source as fits tables.   The following four files are provided for each of COSMOS and XMM-LSS.

\texttt{<Field>\_pybdsf.fits}: contains the pyBDSF total intensity source list for all sources within 0.5$^{\circ}$ of the <field> image centre.  Source parameter columns are as described in the pybdsf documentation\footnote{\url{https://pybdsf.readthedocs.io/en/latest/}}.

\texttt{<Field>\_allsources.fits}: contains spectral and Faraday synthesis results for all sources. The columns are: 

\begin{enumerate}
    \item[1.] the source ID cross-referenced to \texttt{<Field>\_pybdsf.fits};
    \item[2.] the Right Ascension and Declination, with errors;
    \item[3.] the integrated flux density from pyBDSF corrected for the primary beam,  with error;
    \item[4.] the in-band peak total intensity spectral index and error 
    \item[5.] the in-band peak total intensity spectral curvature term and error
    \item[6.] the raw and bias-corrected polarized intensity of the strongest component in the on-source fractional polarisation Faraday Depth spectrum;
    \item[7.] the $P$, $Q$, $U$ fractional polarization of the strongest component in the on-source fractional polarization Faraday Depth spectrum;
    \item[8.] the Faraday depth of the strongest component in the on-source fractional polarization RM synthesis spectrum, with error.
    \item[9.] polarised intensity and Faraday Depth of the peak of the off-source RM synthesis spectrum;
    \item[10.] the effective frequency of the Faraday synthesis.
    \item[11.] 'best\_z\_mightee': The best redshift estimate.
    \item[12.] 'best\_z\_mightee\_err': Photometric redshift uncertainty.
    \item[13.] 'best\_z\_mightee\_origin': Descriptor for the
    origin of the redshift estimate. 'phot\_z' and 'spec\_z' for photometric and spectroscopic redshift from \citet{Adams2021} and \citet{hatfield_hybrid_2022}; 'ZBEST' for spectroscopic redshifts from \citet{Vaccari2022}.
\end{enumerate}

\texttt{<Field>\_poldetections.fits}: contains spectro-polarimetric parameters for all sources with detected polarized emission. Columns are:

\begin{enumerate}
    \item[1.] the source ID cross-referenced to the original pyBDSF catalogue;
    \item[2.] the RA and DEC coordinates with error from the pyBDSF catalogue.
    \item[3.]the integrated flux density in the MFS image from pyBDSF corrected for the primary beam  with error;
    \item[4.] the in-band peak intensity spectral index ($\alpha$) and error;
    \item[5.] the polarization index ($\beta$) with error
    \item[6.] the raw and bias-corrected polarized intensity of the strongest component in the on-source fractional polarisation Faraday Depth spectrum;
    \item[7.] the $P$, $Q$, $U$ fractional polarization of the strongest component in the on-source fractional polarization Faraday Depth spectrum;
    \item[8.] the median fractional Stokes $V$ amplitude across the band,
    \item[9.] the Faraday depth of the strongest component in the on-source fractional polarization RM synthesis spectrum with error;
    \item[10.] the effective frequency of the weighted fractional polarisation RM synthesis spectrum (see Eqn.~\ref{eqn:weights}); 
    \item[11.] the maximum-a-posteriori (MAP) expectation RM from QU-fitting, using the uniform priors described above;
    \item[12.] the 1\,$\sigma$ uncertainty (upper bound) on the MAP RM from QU-fitting derived from the posterior distribution;
    \item[13.] the 1\,$\sigma$ uncertainty (lower bound) on the MAP RM from QU-fitting derived from the posterior distribution;
    \item[14.] 'best\_z\_mightee': The best redshift estimate.
    \item[15.] 'best\_z\_mightee\_err': Photometric redshift uncertainty.
    \item[16.] 'best\_z\_mightee\_origin': Descriptor for the
    origin of the redshift estimate. 'phot\_z' and 'spec\_z' for photometric and spectroscopic redshift from \citet{Adams2021} and \citet{hatfield_hybrid_2022}; 'ZBEST' for spectroscopic redshifts from \citet{Vaccari2022}.
\end{enumerate}

{\texttt{<Field>\_spectra.fits}: contains primary beam corrected Stokes spectra for all sources. Columns are:

\begin{enumerate}
    \item[1.] Source ID
    \item[2.] Frequencies of the channels 
    \item[3.] The primary beam correction for each channel
    \item[4.] Stokes $I$ intensity spectrum 
    \item[5.] The Stokes $I$ intensity model spectrum  
    \item[6.] The Stokes $Q$ intensity spectrum 
    \item[7.] The Stokes $U$ intensity spectrum 
    \item[8.] The Stokes $Q$ background intensity spectrum ($Q_{\rm bkgd}$)
    \item[9.] The Stokes $U$ background intensity spectrum ($U_{\rm bkgd}$)
    \item[10.] The per channel average $Q$, $U$ noise ($\sigma_{QU}$)
\end{enumerate}

Because the number of frequency channels differs among the three
XMM-LSS pointings,  individual \texttt{<Field>\_spectra.fits}
files are provided for each of XMMLSS\_12, XMMLSS\_13 and XMMLSS\_14.

\subsection{Astrometry and Flux Scale}
\subsubsection{Astrometry}

The astrometric offsets were determined by comparing the catalog source positions  with source positions from catalog of compiled optical and infrared
data by \cite{Adams2021}.  We find significant systematic offsets in the 
radio coordinates in all fields. The offsets are the same within errors for all observations of a particular field 
but
different for XMM-LSS and COSMOS, which
suggests the offsets are due to small systematic
errors in the array geometry.
The visibility data from the individual observations of each field were thus combined in uncorrected form and the astrometric corrections per field were derived from the images of the combined visibilities. 
Table~\ref{tab:astrometry} lists the astrometric shifts for each field.  These are
the median difference in RA and DEC for all sources with a match in \cite{Adams2021}.

\begin{table}
    \centering
    \begin{tabular}{|c|c|c|}
    \hline
          Pointing & $\Delta$RA & $\Delta$DEC \\
          & (arcsec) & (arcsec)   \\
    \hline
          XMMLSS\_12 &  0.2743 $\pm$ 0.0008   & 0.5261 $\pm$ 0.0008 \\
          XMMLSS\_13 &  0.2566 $\pm$ 0.0008 & 0.5705 $\pm$ 0.0009\\
          XMMLSS\_14 & 0.2547 $\pm$ 0.0008 & 0.5725 $\pm$ 0.0009 \\
          COSMOS & 0.2657 $\pm$ 0.0010 &  0.1420 $\pm$ 0.0010 \\
    \hline
    \end{tabular}
    \caption{Systematic astrometric offsets of the MIGHTEE-pol early science fields.
    All images and catalogues have been corrected to remove these offsets.}
    \label{tab:astrometry}
\end{table}

The image and catalogue data sets have been corrected for these astrometric errors.
An example bull's eye plot, for the XMMLSS\_14 field, after correction is shown in  Figure~\ref{fig:astrometry}.

\begin{figure}
    \centering
    \includegraphics[width=0.95\columnwidth]{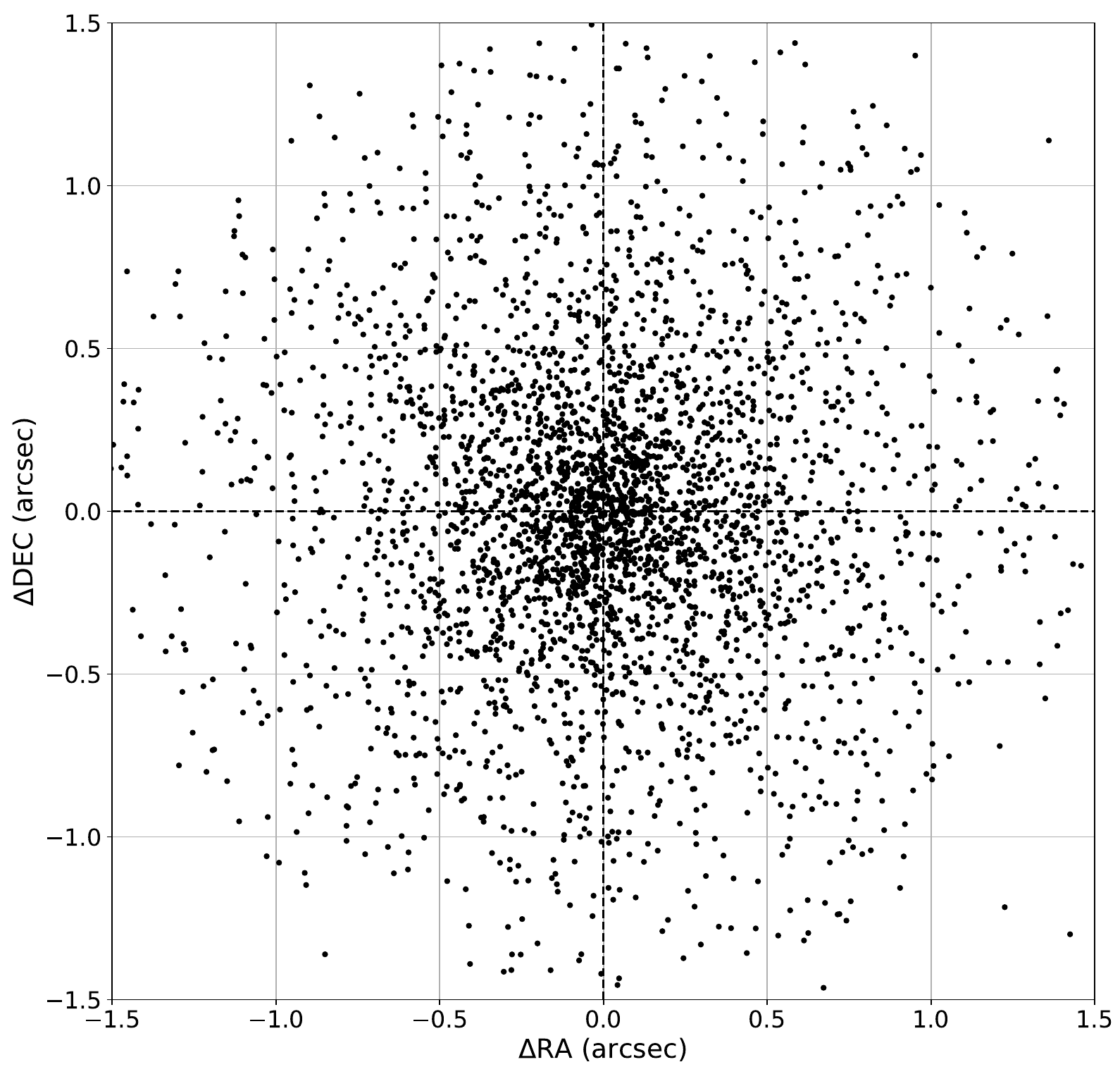}
    \caption{Position differences in RA and DEC for sources in XMM-LSS against the optical positions of \citet{Adams2021} after correcting for the systematic astrometric error in Table~\ref{tab:astrometry}.}
    \label{fig:astrometry}
\end{figure}

\subsubsection{Flux Scale}
\label{sec:fluxscale}

The flux scale was examined by comparing the catalog source fluxes with the fluxes from the MIGHTEE Continuum source catalogs of the same fields \citep{Heywood_2022}. The Heywood et al.\ images have centre frequency at 1.284 GHz, essentially the same
as our MIGHTEE-pol images. We find an overall difference of about 7\% in the source integrated flux densities for the XMMLSS fields, and a difference of 10\% for the COSMOS pointing. In both cases, the flux
densities in MIGHTEE-Pol are lower than the corresponding MIGHTEE-Cont measurements. Figure~\ref{fig:mightee-cont-compare} shows the flux-flux plot of MIGHTEE-Pol integrated flux density vs. MIGHTEE-Cont.

The MIGHTEE Continuum catalogue for the comparison were generated at Briggs robust $= 0.0$, whereas the MIGHTEE-polarization images are generated at robust $= -0.5$. The lower robust for the MIGHTEE-continuum leads to a larger synthesized beam and a higher brightness sensitivity.  The
MIGHTEE-continuum images also include visibilities at long baselines over
the frequency channels that are removed in
our data.  The MIGHTEE-continuum thus uses about 800 MHz of bandwidth, compared to 585\,MHz for our data. 

While we do not expect these effects to cause as much as a 10\% change in measured source flux, these plus differences in the calibration and imaging processes that may explain the offset. We have elected not to apply any correction for the overall flux scaling.  Fractional polarization and Faraday synthesis results are not affected by systematics in flux scale.  The catalogues of source polarization in this paper will 
be accurate to within the limitations of leakage as discussed in section~\ref{sec:leakage}.

\begin{figure}
    \centering
    \includegraphics[width=\linewidth]{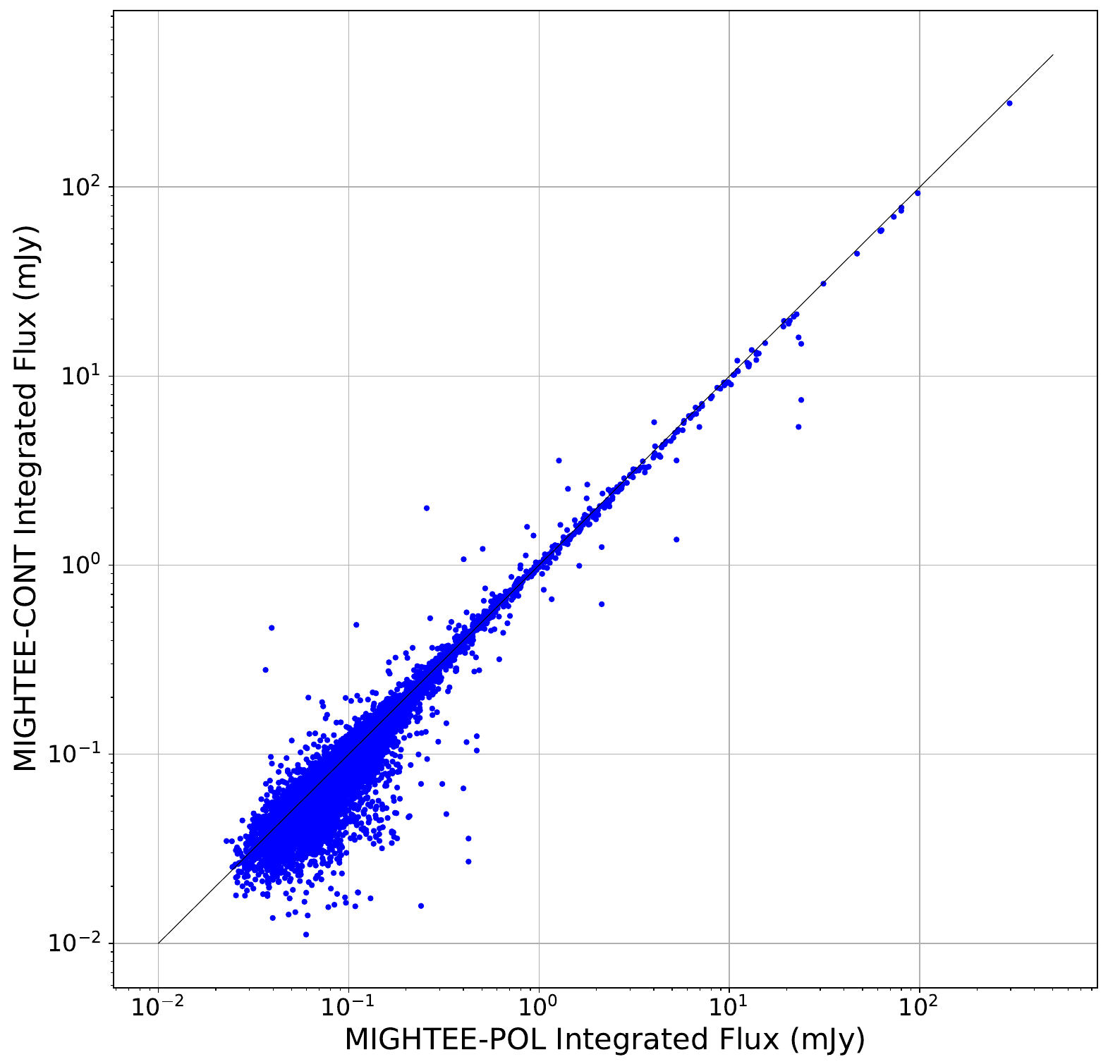}
    \caption{Flux comparison between these data and the MIGHTEE-Cont continuum data release \citep{Heywood_2022} over all the XMMLSS fields. The solid line is a linear fit to the data. We see 6.5\% lower flux density in this case, and $\sim$ 10\% in the case of COSMOS. The exact reason for this flux scale discrepancy is unknown at the moment, so we have elected to not correct our flux density scales to match MIGHTEE-cont.}
    \label{fig:mightee-cont-compare}
\end{figure}


\section{Data Availability}
The raw MeerKAT visibilities for which any proprietary period has expired can be obtained from the SARAO archive at \url{https://archive.sarao.ac.za}. 
The catalogues and images are available at the MeerKAT science archive at
\url{https://doi.org/10.48479/tedr-jf87},
and on the public data repository of the IDIA science gateway at
\url{gateway.idia.ac.za}.

\section*{Acknowledgements}
The MeerKAT telescope is operated by the South African Radio Astronomy Observatory, which is a facility of the National Research Foundation, an agency of the Department of Science and Innovation.
This work was carried out using the data processing pipelines developed at the Inter-University Institute for Data Intensive Astronomy (IDIA) and available at \url{https://idia-pipelines.github.io}. IDIA is a partnership of the University of Cape Town, the University of Pretoria, the University of the Western Cape.
We acknowledge the use of the ilifu cloud computing facility – \url{www.ilifu.ac.za}, a partnership between the University of Cape Town, the University of the Western Cape, the University of Stellenbosch, Sol Plaatje University, the Cape Peninsula University of Technology and the South African Radio Astronomy Observatory. The Ilifu facility is supported by contributions from the Inter-University Institute for Data Intensive Astronomy, the Computational Biology division at UCT and the Data Intensive Research Initiative of South Africa (DIRISA).


\bibliographystyle{mnras}
\bibliography{main} 

\begin{thebibliography}{}
\makeatletter
\relax
\def\mn@urlcharsother{\let\do\@makeother \do\$\do\&\do\#\do\^\do\_\do\%\do\~}
\def\mn@doi{\begingroup\mn@urlcharsother \@ifnextchar [ {\mn@doi@}
  {\mn@doi@[]}}
\def\mn@doi@[#1]#2{\def\@tempa{#1}\ifx\@tempa\@empty \href
  {http://dx.doi.org/#2} {doi:#2}\else \href {http://dx.doi.org/#2} {#1}\fi
  \endgroup}
\def\mn@eprint#1#2{\mn@eprint@#1:#2::\@nil}
\def\mn@eprint@arXiv#1{\href {http://arxiv.org/abs/#1} {{\tt arXiv:#1}}}
\def\mn@eprint@dblp#1{\href {http://dblp.uni-trier.de/rec/bibtex/#1.xml}
  {dblp:#1}}
\def\mn@eprint@#1:#2:#3:#4\@nil{\def\@tempa {#1}\def\@tempb {#2}\def\@tempc
  {#3}\ifx \@tempc \@empty \let \@tempc \@tempb \let \@tempb \@tempa \fi \ifx
  \@tempb \@empty \def\@tempb {arXiv}\fi \@ifundefined
  {mn@eprint@\@tempb}{\@tempb:\@tempc}{\expandafter \expandafter \csname
  mn@eprint@\@tempb\endcsname \expandafter{\@tempc}}}

\bibitem[\protect\citeauthoryear{{Adams}, {Bowler}, {Jarvis}, {H{\"a}u{\ss}ler}
   \& {Lagos}}{{Adams} et~al.}{2021}]{Adams2021}
{Adams} N.~J.,  {Bowler} R.~A.~A.,  {Jarvis} M.~J.,  {H{\"a}u{\ss}ler} B.,
  {Lagos} C.~D.~P.,  2021, \mn@doi [\mnras] {10.1093/mnras/stab1956}, \href
  {https://ui.adsabs.harvard.edu/abs/2021MNRAS.506.4933A} {506, 4933}

\bibitem[\protect\citeauthoryear{Asad et~al.,}{Asad et~al.}{2021}]{asad_2021}
Asad K. M.~B.,  et~al., 2021, \mn@doi [\mnras] {10.1093/mnras/stab104}, 502,
  2970

\bibitem[\protect\citeauthoryear{Bernet, Miniati  \& Lilly}{Bernet
  et~al.}{2012}]{Bernet_2012}
Bernet M.~L.,  Miniati F.,   Lilly S.~J.,  2012, \mn@doi [\apj]
  {10.1088/0004-637X/761/2/144}, 761, 144

\bibitem[\protect\citeauthoryear{{Brentjens} \& {de Bruyn}}{{Brentjens} \& {de
  Bruyn}}{2005}]{bb05}
{Brentjens} M.~A.,  {de Bruyn} A.~G.,  2005, \mn@doi [\aap]
  {10.1051/0004-6361:20052990}, \href
  {https://ui.adsabs.harvard.edu/abs/2005A&A...441.1217B} {441, 1217}

\bibitem[\protect\citeauthoryear{Briggs}{Briggs}{1995}]{briggs1995high}
Briggs D.,  1995, in American Astronomical Society Meeting Abstracts. pp
  112--02

\bibitem[\protect\citeauthoryear{Duncan}{Duncan}{2022}]{Duncan2022}
Duncan K.~J.,  2022, \mn@doi [\mnras] {10.1093/mnras/stac608}, 512, 3662

\bibitem[\protect\citeauthoryear{{Farnes}, {Gaensler}  \& {Carretti}}{{Farnes}
  et~al.}{2014}]{Farnes_2014}
{Farnes} J.~S.,  {Gaensler} B.~M.,   {Carretti} E.,  2014, \mn@doi [\apjs]
  {10.1088/0067-0049/212/1/15}, \href
  {https://ui.adsabs.harvard.edu/abs/2014ApJS..212...15F} {212, 15}

\bibitem[\protect\citeauthoryear{{George}, {Stil}  \& {Keller}}{{George}
  et~al.}{2012}]{George_2012}
{George} S.~J.,  {Stil} J.~M.,   {Keller} B.~W.,  2012, \mn@doi [\pasa]
  {10.1071/AS11027}, \href
  {https://ui.adsabs.harvard.edu/abs/2012PASA...29..214G} {29, 214}

\bibitem[\protect\citeauthoryear{Hales}{Hales}{2017}]{Hales_2017}
Hales C.~A.,  2017, \mn@doi [\aj] {10.3847/1538-3881/aa7aef}, 154, 54

\bibitem[\protect\citeauthoryear{Hammond, Robishaw  \& Gaensler}{Hammond
  et~al.}{2013}]{hammond_2013}
Hammond A.~M.,  Robishaw T.,   Gaensler B.~M.,  2013, A New Catalog of Faraday
  Rotation Measures and Redshifts for Extragalactic Radio Sources (\mn@eprint
  {arXiv} {1209.1438})

\bibitem[\protect\citeauthoryear{Hatfield, Jarvis, Adams, Bowler, Häußler  \&
  Duncan}{Hatfield et~al.}{2022}]{hatfield_hybrid_2022}
Hatfield P.~W.,  Jarvis M.~J.,  Adams N.,  Bowler R. A.~A.,  Häußler B.,
  Duncan K.~J.,  2022, \mn@doi [\mnras] {10.1093/mnras/stac1042}, 513, 3719

\bibitem[\protect\citeauthoryear{Heald et~al.,}{Heald
  et~al.}{2020}]{Heald_2020}
Heald G.,  et~al., 2020, \mn@doi [Galaxies] {10.3390/galaxies8030053}, 8

\bibitem[\protect\citeauthoryear{{Heywood} et~al.,}{{Heywood}
  et~al.}{2022}]{Heywood_2022}
{Heywood} I.,  et~al., 2022, \mn@doi [\mnras] {10.1093/mnras/stab3021}, \href
  {https://ui.adsabs.harvard.edu/abs/2022MNRAS.509.2150H} {509, 2150}

\bibitem[\protect\citeauthoryear{{Hutschenreuter} et~al.,}{{Hutschenreuter}
  et~al.}{2022}]{hutschenreuter2022}
{Hutschenreuter} S.,  et~al., 2022, \mn@doi [\aap]
  {10.1051/0004-6361/202140486}, \href
  {https://ui.adsabs.harvard.edu/abs/2022A&A...657A..43H} {657, A43}

\bibitem[\protect\citeauthoryear{{Jarvis} et~al.,}{{Jarvis}
  et~al.}{2016}]{Jarvis_2016}
{Jarvis} M.,  et~al., 2016, in MeerKAT Science: On the Pathway to the SKA. p.~6
  (\mn@eprint {arXiv} {1709.01901}), \mn@doi{10.22323/1.277.0006}

\bibitem[\protect\citeauthoryear{{Jonas} \& {MeerKAT Team}}{{Jonas} \& {MeerKAT
  Team}}{2016}]{Jonas_2016}
{Jonas} J.,  {MeerKAT Team} 2016, in MeerKAT Science: On the Pathway to the
  SKA. p.~1, \mn@doi{10.22323/1.277.0001}

\bibitem[\protect\citeauthoryear{Mauch \& Sadler}{Mauch \&
  Sadler}{2007}]{Mauch2007}
Mauch T.,  Sadler E.~M.,  2007, \mn@doi [\mnras]
  {10.1111/j.1365-2966.2006.11353.x}, 375, 931

\bibitem[\protect\citeauthoryear{Mauch et~al.,}{Mauch
  et~al.}{2020}]{Mauch_2020}
Mauch T.,  et~al., 2020, \mn@doi [\apj] {10.3847/1538-4357/ab5d2d}, 888, 61

\bibitem[\protect\citeauthoryear{{Pomakov} et~al.,}{{Pomakov}
  et~al.}{2022}]{Pomakov_2022}
{Pomakov} V.~P.,  et~al., 2022, \mn@doi [\mnras] {10.1093/mnras/stac1805},
  \href {https://ui.adsabs.harvard.edu/abs/2022MNRAS.515..256P} {515, 256}

\bibitem[\protect\citeauthoryear{Rau \& Cornwell}{Rau \&
  Cornwell}{2011}]{rau2011multi}
Rau U.,  Cornwell T.~J.,  2011, \aap, 532, A71

\bibitem[\protect\citeauthoryear{Sekhar, Jagannathan, Kirk, Bhatnagar  \&
  Taylor}{Sekhar et~al.}{2022}]{Sekhar2022}
Sekhar S.,  Jagannathan P.,  Kirk B.,  Bhatnagar S.,   Taylor R.,  2022,
  \mn@doi [\apj] {10.3847/1538-3881/ac41c4}, 163, 87

\bibitem[\protect\citeauthoryear{{Shimwell} et~al.,}{{Shimwell}
  et~al.}{2022}]{Shimwell_2022}
{Shimwell} T.~W.,  et~al., 2022, \mn@doi [\aap] {10.1051/0004-6361/202142484},
  \href {https://ui.adsabs.harvard.edu/abs/2022A&A...659A...1S} {659, A1}

\bibitem[\protect\citeauthoryear{Taylor, Stil  \& Sunstrum}{Taylor
  et~al.}{2009}]{Taylor_2009}
Taylor A.~R.,  Stil J.~M.,   Sunstrum C.,  2009, \mn@doi [\apj]
  {10.1088/0004-637X/702/2/1230}, 702, 1230

\bibitem[\protect\citeauthoryear{Taylor et~al.,}{Taylor
  et~al.}{2015}]{Taylor_2015}
Taylor R.,  et~al., 2015, \mn@doi [PoS] {10.22323/1.215.0113}, AASKA14, 113

\bibitem[\protect\citeauthoryear{Vaccari}{Vaccari}{2022}]{Vaccari2022}
Vaccari M.,  2022, {The Spitzer Spectroscopic Data Fusion - Merged
  Spectroscopic Redshift Catalogs in Spitzer Fields},
  \mn@doi{10.5281/zenodo.6368348}, \url
  {https://doi.org/10.5281/zenodo.6368348}

\makeatother
\end{thebibliography}




\appendix

\section{Faraday Completeness}
\label{app:completeness}

We assess the completeness of the catalogued polarised source detections in Faraday depth using simulations. For 100 randomly selected lines of sight from each field, we use the recorded frequency channels to simulate the expected Stokes~Q and U for a population of Faraday thin sources with a range of signal-to-noise ratios (SNR), where the SNR is determined by the individual per channel noise values. The Faraday depth and intrinsic polarisation angle for each simulated source are drawn from uniform distributions with $\phi_0 \sim \mathcal{U}(-200,+200)$ and $\chi_0 \sim \mathcal{U}(-\pi/2,+\pi/2)$, respectively. Completeness (and false-positive rate; FPR) is estimated based on the number of peaks detected in the absolute polarisation as a function of Faraday depth above a level of $4\,\sigma$ (see Section~\ref{sec:pol_detections}), where $\sigma$ is the rms noise in Faraday depth. Detections are considered to be true positives (TP) when the recovered Faraday depth of the peak lies within one FWHM of the RMTF centred on the input Faraday depth value; detections are considered to be false positives (FP) when their peak lies outside this range and has an absolute polarisation greater than that of any true positive present in the data (or greater than the detection threshold in the case that no true positive is detected). Faraday completeness as a function of SNR is shown for each field in Figure~\ref{tab:fcompleteness} and summarised in Table~\ref{tab:fcompleteness}. These results suggest that the polarisation detections are complete above 7\,$\sigma$ for all fields, equivalent to a flux density completeness limit of approximately $17.5\,\mu$Jy\,beam$^{-1}$.
\begin{table}
    \centering
    \caption{Faraday completeness [\%].}
    \label{tab:fcompleteness}
    \begin{tabular}{ccccccccc}
    \hline
     &  & \multicolumn{7}{c}{\textbf{SNR}} \\
     \cline{3-9}\\
    \textbf{Field}  & $\sigma_{\phi}$ & 4 & 5 & 6 & 7 & 8 & 9 & 10 \\\hline
    XMM12     & 2.31 & 61.4 & 89.1 & 98.9 & 100 & 100 & 100 & 100\\
    XMM13     & 2.53 & 61.4 & 89.8 & 98.8 & 99.9 & 100 & 100 & 100\\
    XMM14     & 2.41 &  61.9 & 89.5 & 99.1 & 99.9 & 100 & 100 & 100\\
    COSMOS    & 2.97 &  59.3 & 91.2 & 99.1 & 100 & 100 & 100 & 100\\\hline
    \end{tabular}
\end{table}
\begin{figure}
    \centering
    \includegraphics[width=0.5\textwidth]{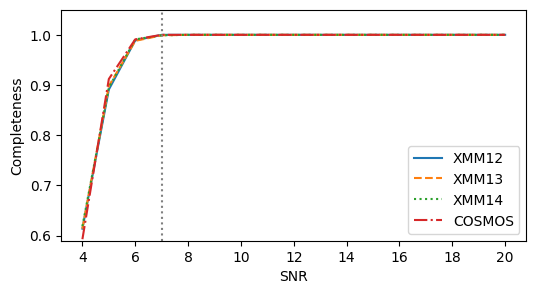}
    \caption{Faraday completeness.}
    \label{fig:fcompleteness}
\end{figure}

\begin{table}
    \centering
    \caption{Faraday False Positive Rate [\%]}
    \label{tab:ffpr}
    \begin{tabular}{cccccccc}
    \hline
     & \multicolumn{7}{c}{\textbf{SNR}}\\
     \cline{2-8}\\
    \textbf{Field}  & 4 & 5 & 6 & 7 & 8 & 9 & 10 \\\hline
    XMM12     & 7.2 & 3.1 & 1.1 & 0.0 & 0.0 & 0.0 & 0.0 \\
    XMM13     & 8.5 & 3.1 & 0.3 & 0.0 & 0.0 & 0.0 & 0.0 \\
    XMM14     &  6.5 & 2.8 & 0.1 & 0.0 & 0.0 & 0.0 & 0.0 \\
    COSMOS    &  6.5 & 2.3 & 0.7 & 0.2 & 0.0 & 0.0 & 0.0 \\\hline
    \end{tabular}
\end{table}

\section{QU Fitting}
\label{app:qufit}

A comparison of the RM values recovered using the peak of the Faraday depth spectrum versus those recovered by fitting directly to the Stokes~Q and Stokes~U data is shown in Figure~\ref{fig:qufitscatter}. It can be seen visually that the recovered values are consistent within 1\,$\sigma$ for all sources. Two sources within the COSMOS field show particularly high uncertainty associated with the RM values recovered from QU-fitting (ID\,1277 \& ID\,2225). The Faraday depth spectra for both sources show significant secondary peaks and it is likely that the contribution of these components to the Stokes~QU data is causing the posterior distributions to become broadened, resulting in larger marginalised uncertainties.

\begin{figure}
    \centering
    \includegraphics[width=\linewidth]{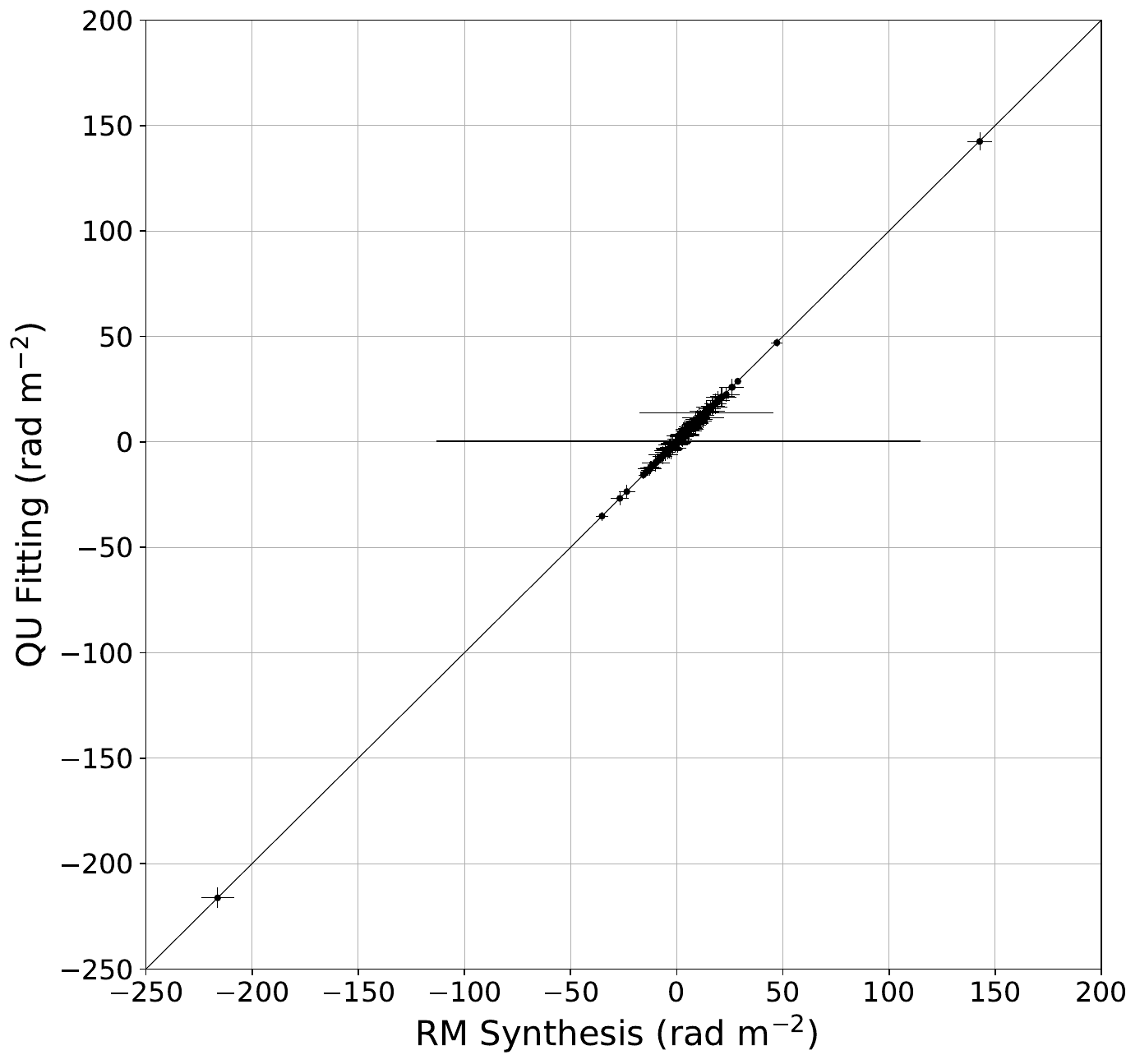}
   \caption{Distribution of catalogue rotation measure (RM) values recovered directly from Faraday depth spectra and those recovered using the QU-fitting method for each field. Error bars show $1\,\sigma$ uncertainties as recorded in the MIGHTEE-POL early release catalogue, see Section~\ref{sec:polarization} for details. \label{fig:qufitscatter}} 
\end{figure}

\bsp	
\label{lastpage}
\end{document}